\documentclass{article}

\usepackage{microtype}
\usepackage{graphicx}
\usepackage{subfigure}
\usepackage{booktabs} 

\usepackage{hyperref}

\usepackage[accepted]{icml2024}

\usepackage{amsmath}
\usepackage{amssymb}
\usepackage{mathtools}
\usepackage{amsthm}

\usepackage[capitalize,noabbrev]{cleveref}

\theoremstyle{plain}

\theoremstyle{definition}

\theoremstyle{remark}

\usepackage[textsize=tiny]{todonotes}

\icmltitlerunning{\emph{convSeq}: Fast and Scalable Method for Detecting Patterns in Spike Data}

\begin{document}

\twocolumn[
\icmltitle{\emph{convSeq}: Fast and Scalable Method for Detecting Patterns in Spike Data}




\begin{icmlauthorlist}
\icmlauthor{Roman Koshkin}{yyy}
\icmlauthor{Tomoki Fukai}{yyy}
\end{icmlauthorlist}

\icmlaffiliation{yyy}{Neural Coding and Brain Computing Unit, Okinawa Institute of Science and Technology, Okinawa, Japan}

\icmlcorrespondingauthor{Roman Koshkin}{roman.koshkin@oist.jp}


\vskip 0.3in
]



\printAffiliationsAndNotice{}  

\begin{abstract}
Spontaneous neural activity, crucial in memory, learning, and spatial navigation, often manifests itself as repetitive spatiotemporal patterns. Despite their importance, analyzing these patterns in large neural recordings remains challenging due to a lack of efficient and scalable detection methods. Addressing this gap, we introduce \emph{convSeq}, an unsupervised method that employs backpropagation for optimizing spatiotemporal filters that effectively identify these neural patterns. Our method's performance is validated on various synthetic data and real neural recordings, revealing spike sequences with unprecedented scalability and efficiency. Significantly surpassing existing methods in speed, \emph{convSeq} sets a new standard for analyzing spontaneous neural activity, potentially advancing our understanding of information processing in neural circuits.
\end{abstract}

\section{Introduction}
\label{intro}

A fundamental property of biological neural networks -- and one that distinguishes them from the majority of modern deep neural networks -- is the ability to change state not only in response to external input, but also spontaneously \citep{arieli1996dynamics,beggs2003neuronal}. A prominent example of this is what is known as "hippocampal replay" \citep{lee2002memory, foster2006reverse, pfeiffer2013hippocampal}, normally observed in 
animals during sleep or periods of immobility, which represents reactivation of spatio-temporal patterns present during a behavioral task performed before. Hippocampal replay has been shown to be crucial for memory, learning and navigation \citep{girardeau2009selective}. In addition, animals' behavior in response to external stimuli depends on the structure of spontaneous activity before and at the time of stimulation \citep{fiser2004small}, which raises the intriguing possibility that spontaneous activity might encode sensory priors and therefore be a form of biological memory. 

To address questions about the role of structured spontaneous activity, a number of methods have been proposed for unsupervised detection of neural activity patterns in the absence of an observable behavioral reference. These existing methods perform well and in reasonable time on modestly sized datasets. However, the study of spontaneous activity would benefit from the analysis of much larger datasets (with hundreds of neurons recorded over several days), which calls for more scalable pattern detection methods.

We introduce an efficient and scalable method for unsupervised detection of spatiotemporal patterns in neural activity based on optimizing a set of constrained 2D filters. Distinct from existing approaches (e.g. \emph{convNMF}, \emph{seqNMF}), we optimize the filters with backpropagation, which allows us to take advantage of popular automatic differentiation frameworks and GPU acceleration. To reduce the number of learnable parameters, we also propose an alternative formulation of our method, in which the filters themselves are parameterized as fixed-width truncated Gaussians. Our speed benchmarks show that the method, which we call \emph{convSeq}\footnote{\texttt{https://github.com/RomanKoshkin/conv-seq}}, works significanly faster than existing pattern detection methods.

Our main contributions are as follows:

\begin{enumerate}
\item Our method advances the SOTA in terms of speed: given the same dataset, it performs over a 100 times faster than similar recently published methods;
\item Unlike \emph{convNMF} and \emph{seqNMF}, which are conceptually similar to our method, ours provides uncertainty estimates for the patterns detected, without requiring multiple optimization runs.
\end{enumerate}

The rest of the paper is structured as follows. Section \ref{sec:relatedwork} briefly reviews existing methods for detecting spatio-temporal patterns in neural data. After introducing our method (in Section \ref{sec:methods}), we showcase its ability to detect various patterns in synthetic and real datasets (in Section \ref{sec:experiments}), as well as accuracy and speed comparisons with a selection of other methods (in Section \ref{sec:benchmarks}). We discuss training, implementation and the choice of hyperparameters in Section 
 \ref{appendix:guidance} and conclude with future directions and final remarks in Section \ref{section:Conclusions}.

\section{Related Work}
\label{sec:relatedwork}
In general, classic methods working under linear assumptions, such as PCA and ICA \citep{JUTTEN19911}, struggle to capture spatiotemporal patterns in neural activity, as they tend to merge them into a single "large component” \citep{peter2016sparse, williams2020point}. This key limitation motivated many previous works which proposed alternative methods for detecting spatiotemporal structure in neural data, without using external reference events. For example, in a departure from traditional distance metrics, \citet{watanabe2019unsupervised} used edit similarity between potential spike patterns to identify cell assembly sequences. Techniques like the “intersection matrix” used by \citet{schrader2008detecting, torre2016asset} specifically targeted synchronous events  (aka synfire chains). Further expanding the analytical toolbox for neural data, \citet{shimazaki2012state} used state-space modeling to detect higher-order spike correlations. More recent advancements include the clustering method based on optimal transport by \citet{grossberger2018unsupervised}, innovative point process models by \citet{williams2020point} and \citet{li2022online} and organizing neural responses along a one-dimensional manifold to expose the patterns \citep{Stringer2023.07.25.550571}.

The convolutional non-negative matrix factorization (\emph{convNMF}) introduced by \citet{smaragdis2004non, smaragdis2006convolutive} and first applied to in-vitro neural data by \citet{peter2016sparse} is conceptually closest to our approach. \emph{convNMF} and its improved derivative, \emph{seqNMF}      \citep{mackevicius2019unsupervised}, aim to jointly estimate both the templates of the recurrent patterns and the time course of their activity. In contrast, our approach, as we describe next, only optimizes the templates (which we call “filters”) and does so using backpropagation.

\section{Methods}
\label{sec:methods}

The input to our model is a binary matrix $\mathbf{X} \in \{1, 0\}^{N\times T}$, which represents a simultaneous recording of $N$ neurons for $T$ time bins (also referred to as “time steps”), such that $X_{n,t}=1$ if there is a spike on the $n$-th neuron in time bin $t$, and $X_{n,t}=0$ otherwise. We seek to find $K$ 2D filters $\mathbf{W}^{(k)} \in \mathbb{R}^{N\times M}$, such that each of them responds preferentially to one of $K$ unknown spatiotemporal patterns (also referred to as spike sequences throughout the paper). Each of the $K$ patterns are repeated \emph{inexactly} – due to variations in the relative timing (jitter) and dropout of spikes – an unknown number of times. The choice of $M$ and $K$ depends on the length (in time steps) and number of the patterns assumed to be present in the data.

\subsection{Formulation with direct filter optimization}
\label{subsection:formulaiton_1}

We first describe how the filters $\mathbf{W}^{(k)}$, $k \in \{1,\dots, K\}$, can be found directly by minimizing the following loss function:

\begin{equation}
\begin{split}
    \mathcal{L}(\mathbf{W}) = \sum_{k=1}^{K} - \mathrm{Var}(\mathbf{\hat{x}}^{(k)}) + \beta_{\mathrm {TV}} \mathrm {TV}(\mathbf{\hat{x}}^{(k)}) + \\ + \beta_{\mathrm {xcor}} \sum_{l > k}^{K} \rho_{\mathbf{\hat{x}}^{(l)} \mathbf{\hat{x}}^{(k)}}[j]
\end{split}
\label{eq:conv1}
\end{equation}

where $\mathbf{\hat{x}}^{(k)} = \mathrm{softmax}(\mathbf{W}^{(k)}) * \mathbf{X}$, and “*” stands for convolution. The convolution is performed with no padding along the dimension of neurons and sufficient zero padding along the time dimension to ensure that $\mathbf{\hat{x}}^{(k)}$ has shape $1 \times T$. Softmax is computed over the time dimension of $\mathbf{W}^{(k)}$. $\mathrm {TV}({\hat{\mathbf{x}}}^{(k)}) = \frac{1}{T}\sum_{t=1}^{T-1}(\hat{x}^{(k)}_t - \hat{x}^{(k)}_{t+1})^2$ and $\sum_{l > k}^{K} \rho_{\mathbf{\hat{x}}^{(l)} \mathbf{\hat{x}}^{(k)}}[j]$ are total variation and cross-correlation over $j$ time steps, respectively. The first term in Eq. \ref{eq:conv1} maximizes the variance of the $k$-th filter's total response to the data. The idea is that if there exists a repeating pattern, the right filter (when convolved with the data) will produce peaks at the times of that pattern's occurrence. Importantly, while each filter's total response stays constant (that is $\sum \big[ \mathrm{softmax}(\mathbf{W}^{(k)}) * \mathbf{X}\big] = \sum \mathbf{X}$), the variance of its total response is maximized when the filter has a good match with some repeating pattern. Keeping the filter's total response constrained makes it easy to bootstrap confidence intervals for the height of peaks in $\mathbf{\hat{x}}^{(k)}$, which can be used for testing the significance of the patterns detected (Section \ref{subsec:stats}). The total variation term helps reduce the filters' response to background activity (i. e. neural activity unrelated to any pattern), reduce the false positive rate, and facilitate visual interpretation of results (Appendix \ref{appendix:total_variation}). Finally, the cross-correlation term in Eq. \ref{eq:conv1} encourages filter diversity when $K > 1$. That is, it prevents the filters from becoming "tuned" to the same (stronger or overrepresented) pattern. The weights of the total variation and cross-correlation penalty terms as well as other hyperparameters are listed in Appendix \ref{HPs}.

\subsection{Visualization of structured spontaneous activity}

\begin{figure}[t]
\centering
\includegraphics[width=0.48\textwidth]{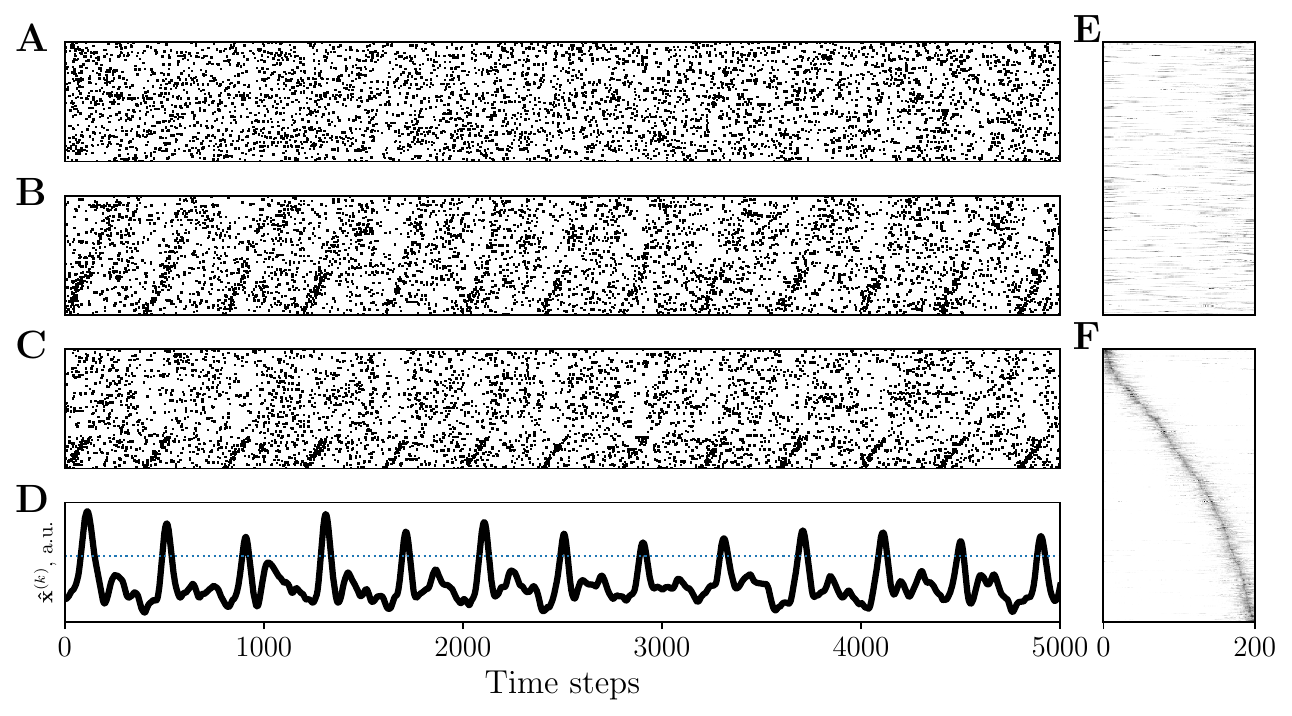}
\vskip -0.1in
\caption{Overview of our method (with $K=1$). Initially, no sequences are visible in the input data matrix $\mathbf{X}$ (A). After model fitting, the convolution $\mathbf {\hat x}^{(k)}$ of the optimized filter $\mathbf{W}^{(k)}$ (E) with the data contains clearly visible peaks (D), suggesting the presence of repeating patterns. The peaks in $\mathbf{\hat x}^{(k)}$ exceeding the significance threshold $\alpha$ (dotted line in D) indicate significant detections. To make the patterns visible, the rows of the optimized filter $\mathbf{W}^{(k)}$ are sorted according to the latency of maximum value (F). Finally, rearranging the rows of $\mathbf X$ with the obtained sorting indices exposes the sequences (B), which closely match the ground truth (C). The y-axis in all the panels except D corresponds to neuron IDs. Panles E and F are not to scale with Panels A-D.}
\label{fig:fig1}
\end{figure}

The presence, strength and temporal location of the patterns is captured by ${\hat{\mathbf{x}}}^{(k)}$: its peaks correspond to the times at which the pattern is expressed in neural activity (Fig. \ref{fig:fig1}D). These peaks alone, however, only suggest the presence of a pattern, and it is desirable to represent the data in a way that makes the detected structure clearly visible (e. g. in hippocampal recordings in which theta sequences are expected). To reveal the patterns, the optimized filters are sorted so that per-row maxima become temporally ordered. Then the sorting indices are used to rearrange the order of neurons in $\mathbf{X}$. We summarize this in Fig. \ref{fig:fig1} and Appendix \ref{appendix:algo1}. We also note that depending on pattern complexity and strength, as well as parameter initialization, variations of the recovered patterns' shape are to be expected.

\subsection{Formulation with parameterized Gaussian filters}
\label{subsection:formulaiton_2}

In the above formulation, we seek to optimize the randomly initialized filters $\textbf{W}^{(k)}$ directly, which means $N \times W \times K$ trainable parameters. However, assuming that patterns are sequences of spikes, whose relative timing is distorted by spike timing jitter, and that this jitter is Gaussian, we can reduce the number of trainable parameters by a factor of $N$. Specifically, at each optimization step, we can parameterize the $n$-th row in the $k$-th filter $\mathbf{W}^{(k)}$ as a truncated Gaussian function $f(\cdot)$ with mean $\mathrm{\mu}_n^{(k)}$ and a fixed value of $\sigma$. In this way, we only need to optimize the means of the Gaussians in each row. In this formulation, the softmax function is no longer needed as the filter's impulse response is now constrained by the Gaussian function truncated to the filter's width $M$: $\mathbf{\hat{x}}^{(k)} = \mathbf{W}^{(k)} * \mathbf{X}$, such that $\mathbf{W}^{(k)}_{n,:} = f(\mu_n^{(k)}, \sigma^2, 1, M)$, and $n \in \{1, \dots, N\}$. While in terms of speed this formulation performs on par with the one described in Section \ref{subsection:formulaiton_1}, it offers a way to steer the model towards specific solutions by incorporating inductive biases into the filter design. For example, it should also be possible to learn per-neuron standard deviations $\sigma_n^{(k)}$ (although at the expense of doubling the number of trainable parameters) to capture each neuron’s temporal jitter and its degree of participation in a pattern, but we leave this question to future work.

\subsection{Statistical testing}
\label{subsec:stats}
\begin{figure}[H]
\centering
\includegraphics[width=0.49\textwidth]{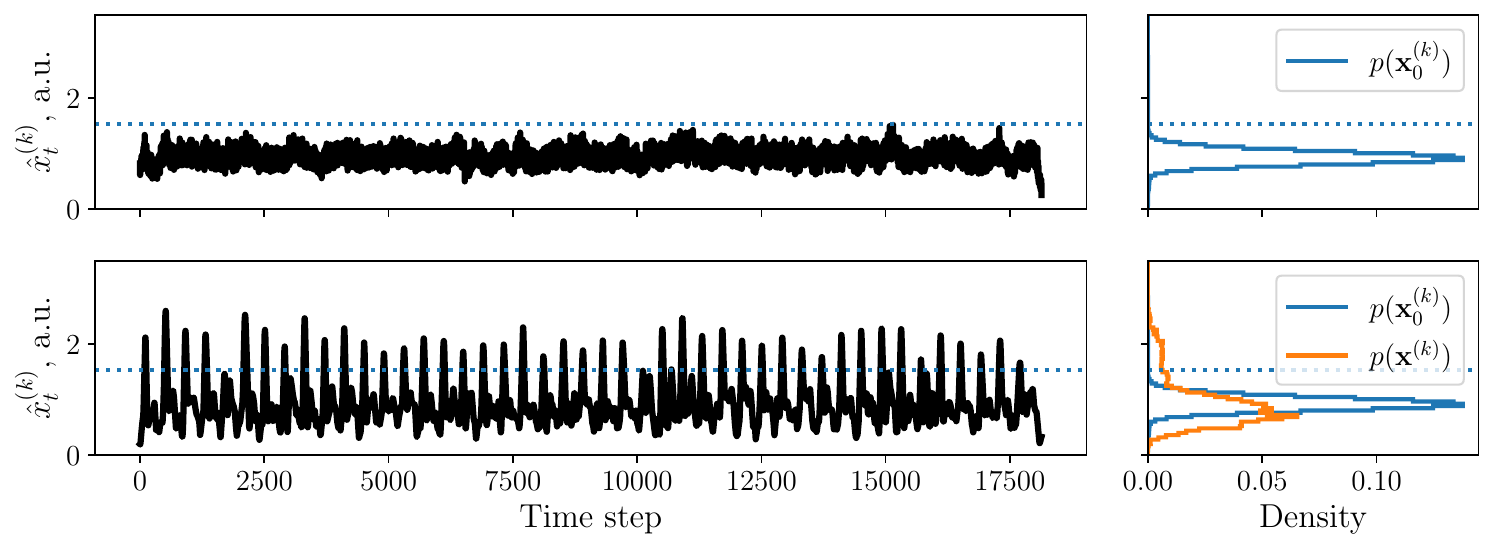}
\vskip -0.1in
\caption{Before optimization (top row) no pattern is detected as the peaks in $\mathbf{\hat{x}}^{(k)}$ lie below $\alpha$ (dotted line). After optimization (bottom row) multiple occurrences of the pattern are detected. The curves in the panels on the right illustrate $\mathbf{\hat{x}}^{(k)}$ as densities. Blue curves illustrate the distribution of values in $\mathbf{\hat{x}}^{(k)}$ expected from a random filter, red curve shows the distribution of responses of the optimized filter.}
\label{fig:fig2}
\end{figure}

We consider a detection of the $k$-th pattern to be statistically significant at some time step $t$ if $\hat{x}_{t}^{(k)} >= \alpha$, where $t\in \{1,\dots,T\}$ and $\alpha$ is a significance criterion, which is determined for each dataset individually. To estimate $\alpha$, we construct 1000 random filters and get $\mathbf{\hat{x}}_0^{(k)}$ = $\mathbf{X} * \mathbf{W}_0^{(k)}, k \in \{1,\dots, 1000\}$, out of which we construct the null distribution $p(\mathbf{\hat{x}}_0^{(k)})$. Computing its mean, $\mu_0$, and standard deviation, $\sigma_0$, we set $\alpha$ to be four\footnote{Empirically, setting $\alpha$ to 4 standard deviations ensures a very low false positive rate.} standard deviations above the mean, i.e. $\alpha = 4 \sigma_0 + \mu_0$ (Fig. \ref{fig:fig2}). Depending on the level of confidence desired, a more lenient threshold can be chosen. Besides significance testing, $\alpha$ can be used for early stopping: for example, optimization can be terminated once a desired number of peaks in $\mathbf{\hat{x}}^{(k)}$ reach or exceed $\alpha$. 

\section{Experiments}
\label{sec:experiments}

Since both formulations of our method perform comparably, here we report the results obtained using the first formulation. For the second formulation please refer to Appendix \ref{appendix:2nd_additional}.

\subsection{Accuracy performance metrics}

To evaluate the model’s accuracy performance we use the following three metrics: \emph{true positive rate} – the proportion of times a sequence is detected by the corresponding filter. A true positive is scored when the $k$-th filter responds with a significant peak in $\mathbf{\hat{x}}^{(k)}$ within no more than $M//2$ time steps of the ground truth label marking the middle of a sequence. This margin of $M//2$ time steps is needed because the response of an optimized filter to its preferred pattern is not guaranteed to coincide perfectly with the middle of the pattern. This is especially the case if a filter's chosen width exceeds the width of the pattern. \emph{False positive rate} – the proportion of times a filter produces a significant peak to the wrong sequence or background activity (that is when no sequence is expressed). Finally, \emph{false negative rate} – when a filter fails to produce a significant peak in response to its corresponding sequence.

\subsection{Synthetic data}

We first test our method on three synthetic datasets. To simulate biologically realistic spike statistics, these datasets were constructed by embedding different spike sequences into a matrix of background activity $\mathbf{X} \in \{0,1\}^{N\times T}$ obtained by permuting the rows and columns of the real mouse CA1 recording described in Sec. \ref{subsec:ca1}. To facilitate comparisons, in all the experiments the shape of the synthetic datasets ($N=452$, $T=18137$) and filters ($N=452$, $M=200$) were kept the same unless indicated otherwise.

\begin{figure}[H]
\centering
\includegraphics[width=0.48\textwidth]{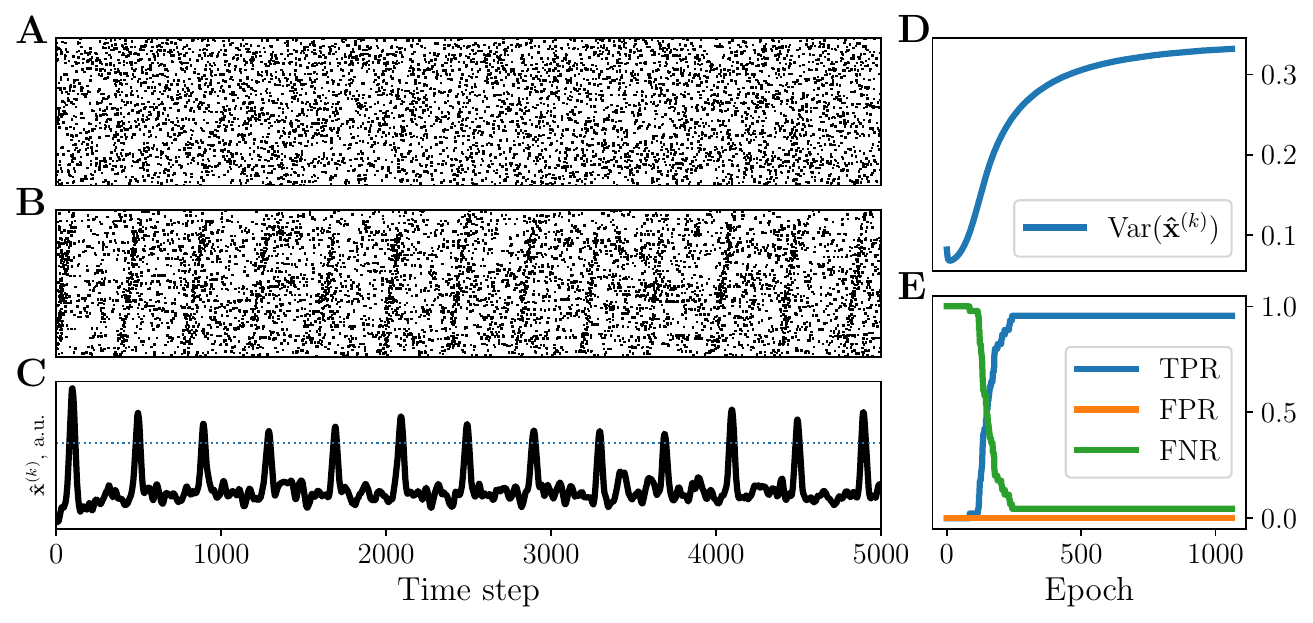}
\vskip -0.1in
\caption{(A) and (B) depict the first 5000 time steps of the data before and after sorting based on the optimized filter, whose convolution with the data is shown in (C). (D) and (E) show evolution of the variance of the filter's response and performance metrics, respectively, over the course of optimization.}
\label{fig:fig3}
\end{figure}

\paragraph{One sequence.} We first consider the simplest case, in which a datasets contains 45, 30 and 22 occurrences of a sequence of one type (i.e. 400, 600 and 800 time steps apart, respectively) consisting of 80 neurons, each of which dropped with a probability of 0.2. We also add a Gaussian temporal jitter with a standard deviation of 10, 20 and 30 time steps. This results in 9 datasets. Fig. \ref{fig:fig3} illustrates the case with 45 sequence repetitions, dropout of 0.2 and a jitter of 10, where our model is able to detect all the sequence occurrences. Fig. 4 summarizes our model's performance on all the 9 datasets, each run 8 times.

\begin{figure}[H]
\centering
\includegraphics[width=0.48\textwidth]{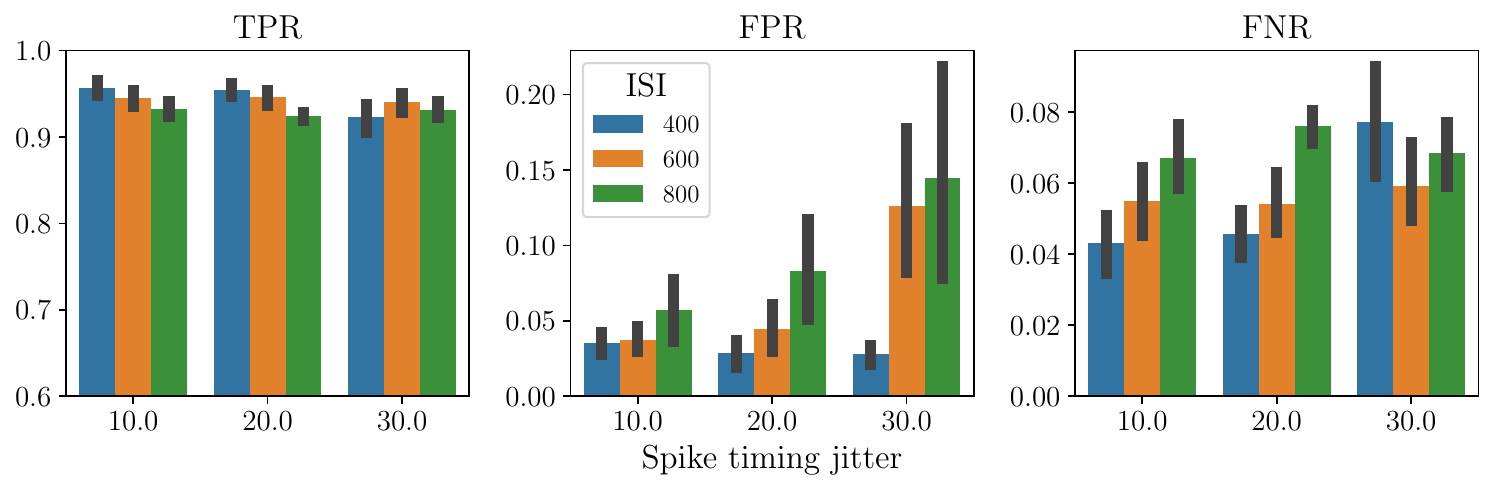}
\vskip -0.1in
\caption{Model's accuracy performance as a function of spike timing jitter and inter-sequence interval (ISI). Error bars indicate the 95\% confidence intervals computed over 8 runs.}
\label{fig:fig4}
\end{figure}

As expected, the accuracy performance degrades as individual spike timings within a sequence occurrence deviate more from their ideal timing (higher spike jitter) and as the sequence occurrences become less frequent (longer inter-sequence interval) (Fig. \ref{fig:fig4}). As we show in Section \ref{sub:detection_perf}, the accuracy performance also depends on how many spikes are dropped from the sequence (spike sequence sparsity), and the number of neurons participating in the sequence (sequence length).

\paragraph{Two sequences overlapping in space.} We next test the ability of our method to detect two partially overlapping sequences. This is a more challenging scenario because the filters will have to compete for the neurons shared by both sequences. We used the same parameters as in Experiment 1, except that instead of 1 sequence of 80 neurons, we embedded 2 sequences of 100 neurons (overlapping by 50 neurons) alternating approximately every 480 time steps. Overall, each sequence was repeated 22 times (44 repetitions in total).

\begin{figure}[H]
\centering
\includegraphics[width=0.48\textwidth]{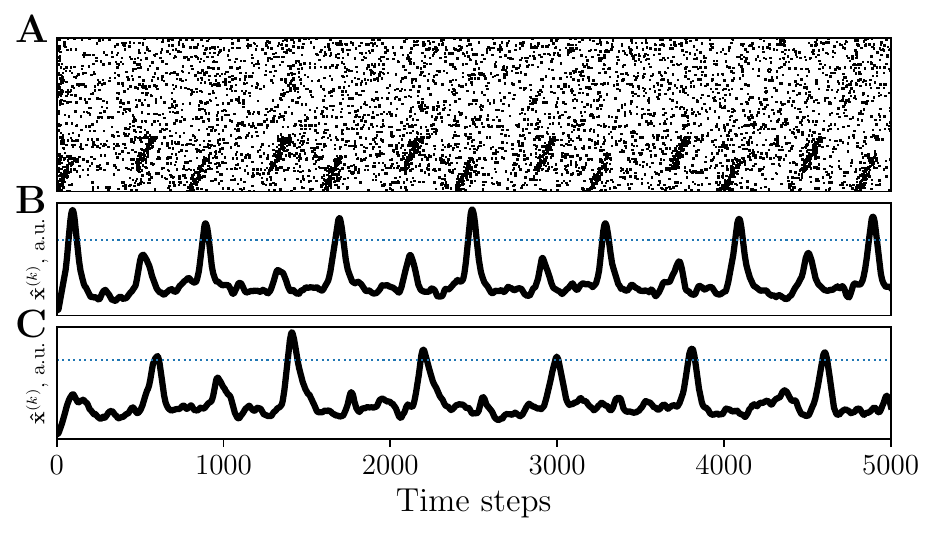}
\vskip -0.1in
\caption{The model can correctly detect all the occurrences of two partially overlapping sequences. (A) fragment of the ground truth (original data before permuting the rows). Response of the first and second filter after optimization are shown in (B) and (C), respectively.}
\label{fig:fig5}
\end{figure}

Despite the partial overlap in space, the model is still able to disentangle all the sequence occurrences correctly (Fig. \ref{fig:fig5}). We note, however, the presence of undesirable peaks in the response of the other pattern (Fig. \ref{fig:fig5}B,C), which indicates that \emph{unidirectional} patterns with shared neurons are hard to disentangle cleanly. Although those undesirable peaks do not reach the threshold of significance, they pose a potential issue for the detection of short or closely adjacent sequences with shared neurons. We leave detailed treatment of such cases as well as further improvements of the method to future work.

\begin{figure}[H]
\centering
\includegraphics[width=0.48\textwidth]{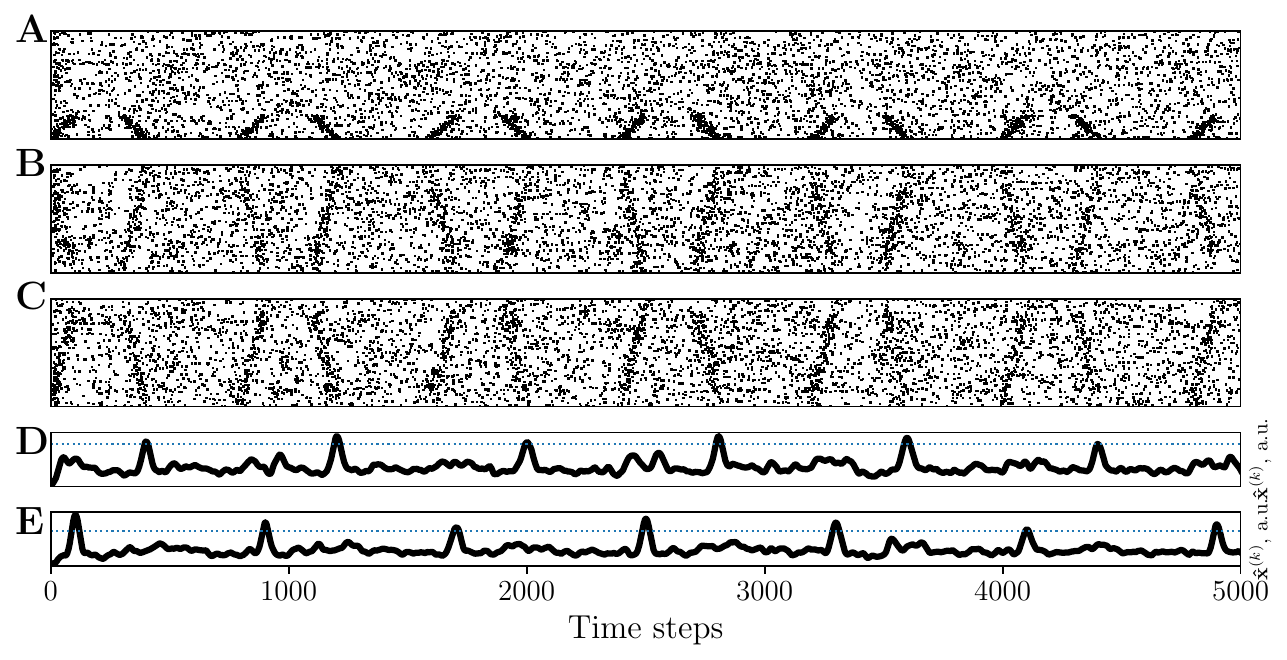}
\vskip -0.1in
\caption{The model successfully detects forward and reverse instances of a bidirectional sequence. For illustration, the original data in (A) is shown before permuting the order of neurons. Sorting the permuted data with the first (B) and second (C) optimized filter exposes the forward and reverse sequences. The lines in (D) and (E) show the first and second filters’ responses, respectively. The dotted horizontal line in (D) and (E) marks the significance threshold.}
\label{fig:fig6}
\end{figure}

\paragraph{Bidirectional sequences with full overlap in space.} Our third synthetic dataset contained two \emph{bidirectional} sequences (i.e. expressed in both forward and reverse order), constructed in the same way as in the previous experiment, but consisting of fully shared 100 neurons (Fig. \ref{fig:fig6}).

\paragraph{Time-warped sequences.}
\label{appendix:time_warp}

\begin{figure}[H]
\centering
\includegraphics[width=0.48\textwidth]{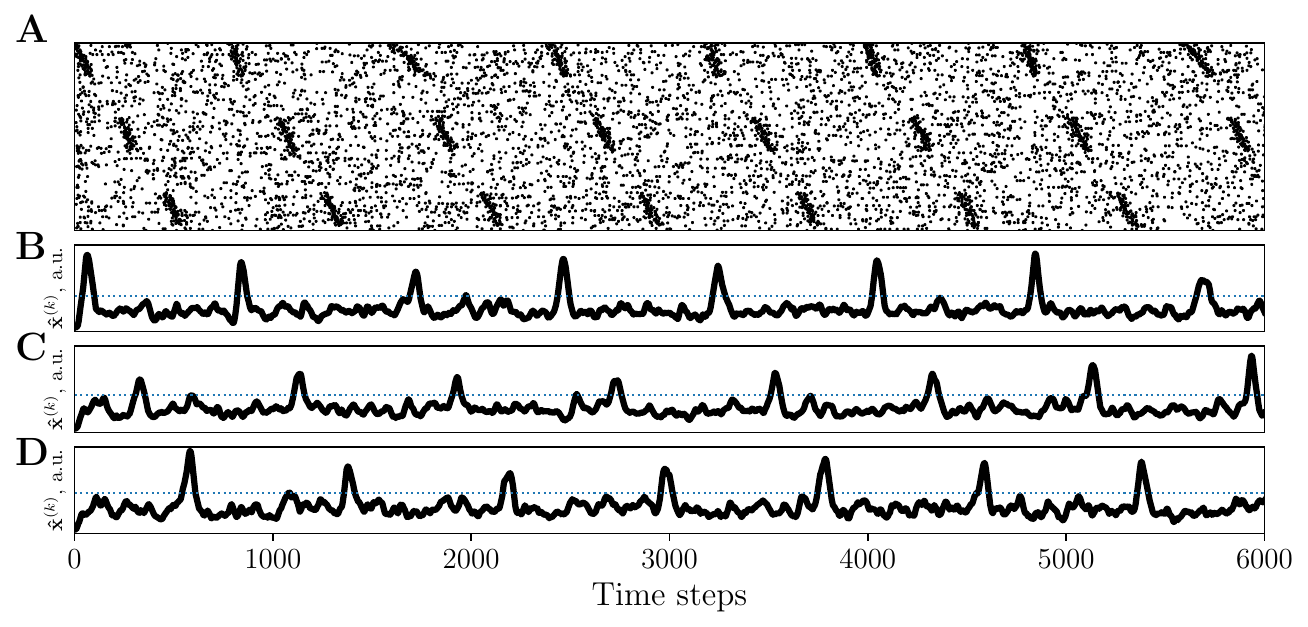}
\vskip -0.1in
\caption{The model detects 3 sequences one of which (top sequence in A) is time-warped with a factor randomly chosen from $\{0.6, 1.0, 1.8, 2.2\}$. Panels B, C and D show $\mathbf{\hat{x}}^{(k)}$, $k \in \{1,2,3\}$.}
\label{fig:time_warp}
\end{figure}

It is possible for the same sequence (that is, a sequence involving the same neurons, and firing in approximately the same order) to be expressed over more than one time scale. Our approach is robust enough to handle mildly time-warped sequences (Fig. \ref{fig:time_warp}).

\begin{figure}[H]
\centering
\includegraphics[width=0.48\textwidth]{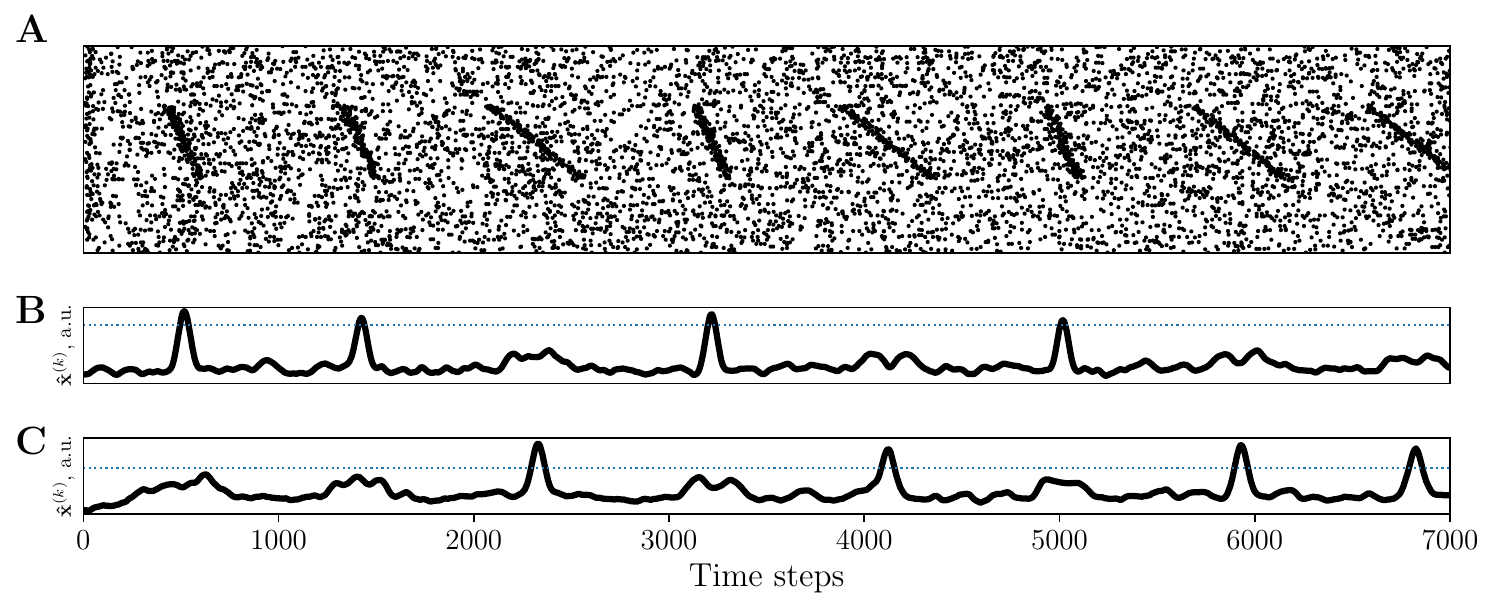}
\vskip -0.1in
\caption{Handling significantly time-warped sequences (A). Filters optimized separately with $M=200$ and with $M=600$ produce a strong detection signal in response to short (B) and long (C) versions of the sequence, respectively. The dotted horizontal line in (B) and (C) marks the significance threshold.}
\label{fig:diff_timescale}
\end{figure}

However, if one expects significant time warping (e. g. when the duration can vary by a factor of more than 2), we recommend running optmizaton with $K = 1$ with several progressively larger values of $M$. Smaller filters should become tuned to more “compressed” versions of the sequence, while larger ones will tend to capture its slower versions. We illustrate this on a synthetic dataset ($N = 452$, dropout of 0.2, spike timing jitter of 15) with one sequence of 160 neurons which unfolds at two different speeds (the second is 3 times slower than the first). We begin by setting $M = 200$ and optimize the first filter. As expected, this filter only responds to the short sequence (Fig. \ref{fig:diff_timescale}), because the longer sequence is not fully contained within the filter’s temporal length. Next, we set $M = 600$ and optimize the second filter. After optimization, this wider filter produces higher peaks in response to the longer sequence (Fig. \ref{fig:diff_timescale}).

\paragraph{Sequences of 2D place cells.}
\label{sub:tmaze}

\begin{figure}[H]
\centering
\includegraphics[width=0.48\textwidth]{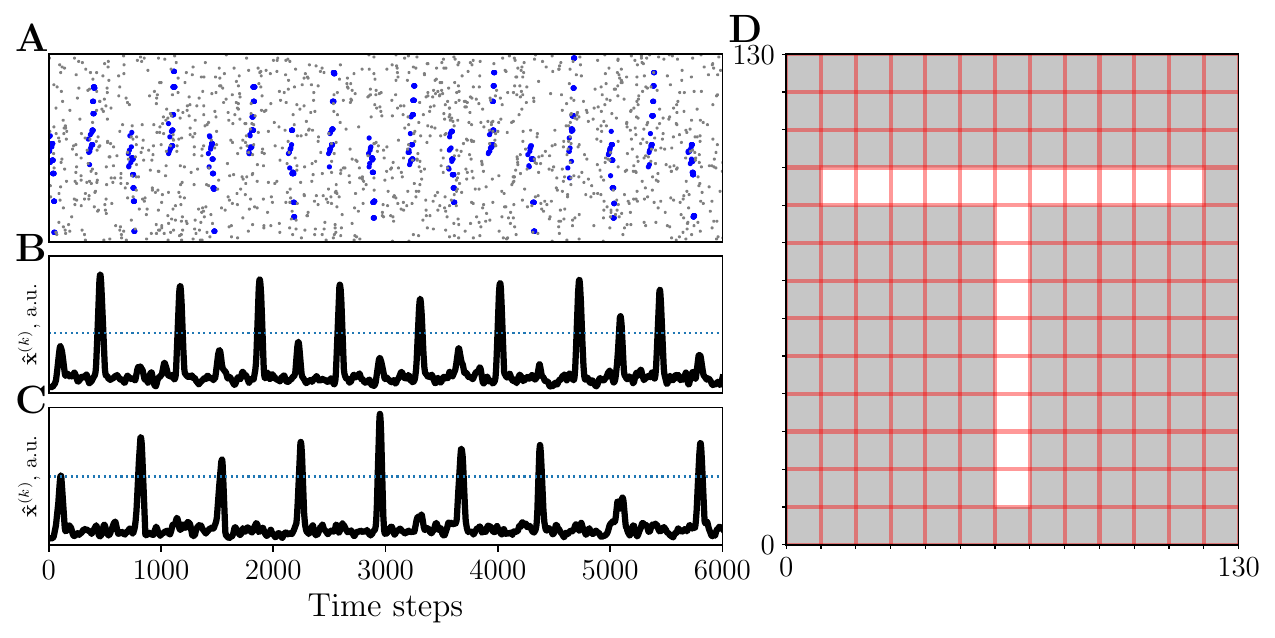}
\vskip -0.1in
\caption{Handling 2 spatially overlapping spike sequences (blue dots in A) of a mouse traversing place fields tiling a $130 \times 130$ cm 2D enclosure (D). Panels B and C show $\mathbf{\hat{x}}^{(k)}$, $k \in \{1,2\}$.}
\label{fig:tmaze}
\end{figure}

Finally, we consider a synthetic dataset simulating the activity of $N=169$ place cells tiling a 1.3 m x 1.3 m enclosure (Fig. \ref{fig:tmaze}D). As the mouse runs in the T-maze inside the enclosure (from the bottom of the vertical arm to either the right or the left end of the horizontal arm, every 300 time steps), the place cells spike with a probability inversely proportional to the distance between the center of their place fields and the animal’s position (2D Gaussian with $\mathbf {\mu} = (x, y)$, $x, y \in \{5, \dots, 125\}$, $\sigma = 10$ cm). For each $t \in \{1, \dots, 6000\}$ we sample one spike. If the spike occurs, for example, on the neuron whose place field is centered at 95 cm up from the bottom and 15 cm from the left wall of the enclosure, we make a square matrix of zeros, set its element [3, 1] to 1, vectorize it and add it to a list. Repeating this process for $T = 6000$ timesteps we obtain a matrix of size $N \times T$, which after adding background noise and jittering the spikes produces the dataset (Fig. \ref{fig:tmaze}A). Notice that the sequences appear broken because the place fields adjacent in 2D are no longer adjacent after vectorization. Fig. \ref{fig:tmaze} (B,C).

\subsection{Real data}

\paragraph{Recording from the CA1 area of the mouse hippocampus.}
\label{subsec:ca1}

Next we tested the ability of our method to expose place cell sequences in real neural data . We used a dataset\footnote{The dataset is available at https://github.com/zivlab/island and represents a binary matrix obtained by thresholding the original $\text{Ca}^{2+}$ imaging data.} from \citet{rubin2019revealing}, which is a recording of CA1 neurons of a mouse running on a linear track and collecting water rewards dispensed at its ends. 

\begin{figure}[H]
\centering
\includegraphics[width=0.48\textwidth]{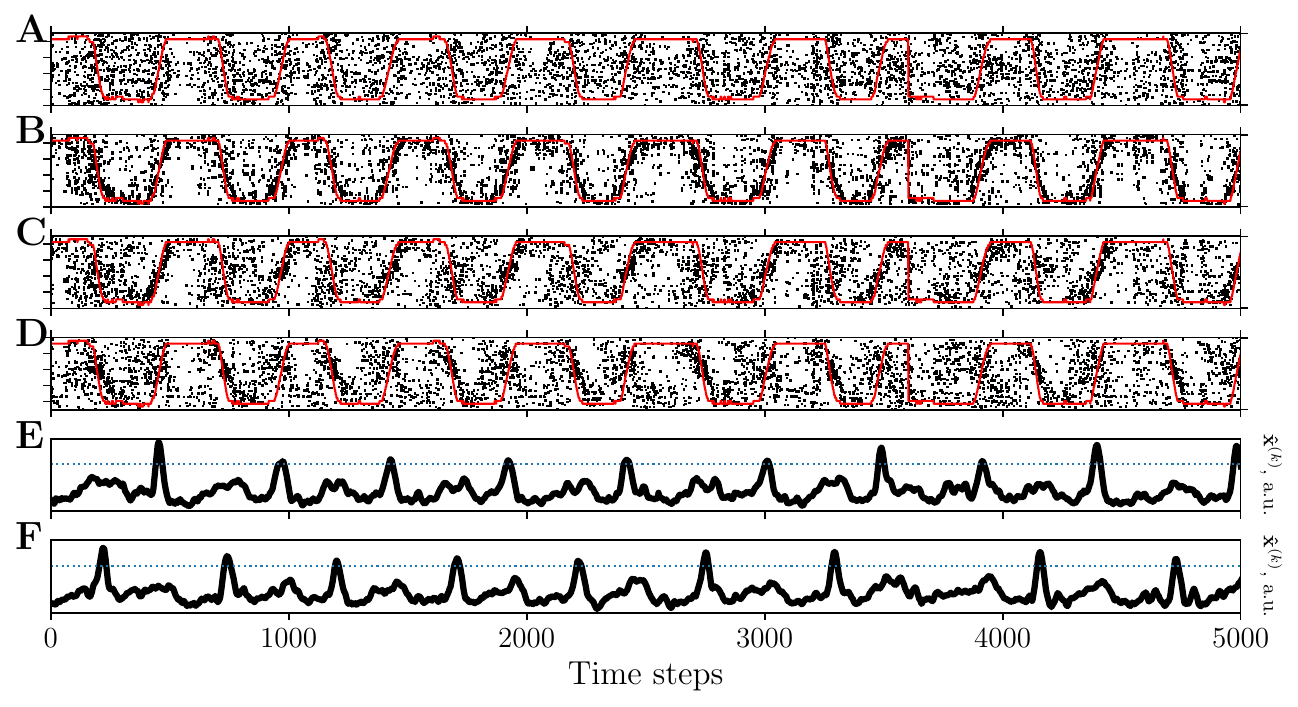}
\vskip -0.1in
\caption{The red line in (A-D) indicates the animal’s position on the track. (B) shows the ground truth, in which neurons of the original dataset (A) are sorted according to their known place fields. In (C) and (D), neurons of the original dataset are rearranged with the indices obtained by sorting the first and second optimized filters, respectively. The first and second filters' responses are shown in (E) and (F), respectively. The dotted horizontal line in (E) and (F) marks the significance threshold.}
\label{fig:fig9}
\end{figure}

In this experiment the position of the mouse was recorded simultaneously with the neural activity, and so we can verify the detected patterns against the ground truth -- the ordering of neurons based on their known place fields. A neuron's place fields are determined by measuring its activity across the environment: the more a neuron fires in a particular location, the more “preferred” that location is. As the animal goes through different locations on the track, neurons with similar place tuning are more likely to spike together, and this information can be used to rearrange the order of neurons to make place cell sequences clearly visible (Fig. \ref{fig:fig9}B).

With $K=2$ and $M=200$, our model was able to disentangle the forward and backward sequences of place cells, with peak activation of the corresponding filters exceeding the significance threshold.

\paragraph{Songbird higher vocal center.}
\label{subsec:hvc}

\begin{figure}[H]
\centering
\includegraphics[width=0.48\textwidth]{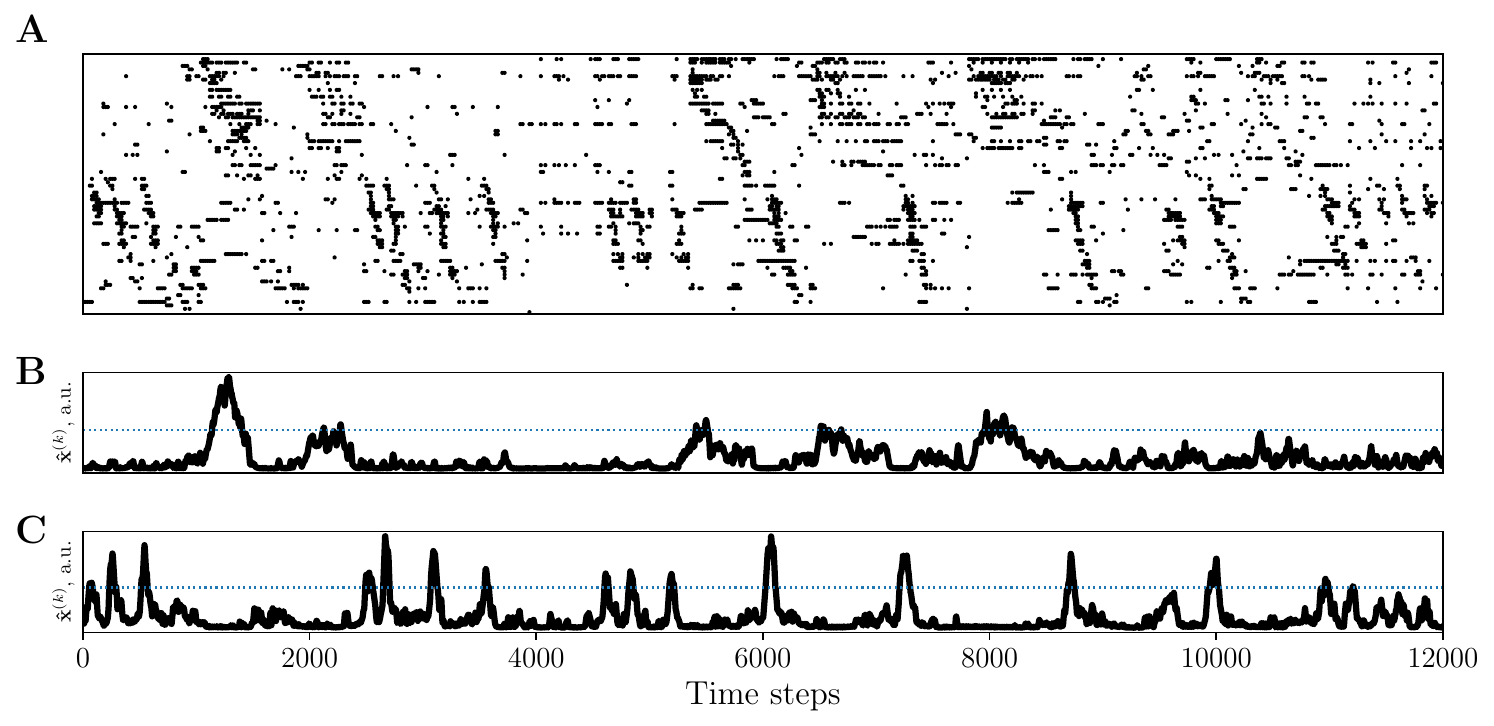}
\vskip -0.1in
\caption{Raw data recorded from the songbird higher vocal center (A). Response of the first
and second filter after optimization are shown in (B) and (C), respectively. The dotted horizontal lines in (B) and (C) marks the significance threshold.}
\label{fig:hvc}
\end{figure}

We also applied our method to a dataset recorded from the higher vocal center (HVC) of a songbird. HVC neurons are known to produce precisely timed sequences, which our method was able to extract (Fig. \ref{fig:hvc}).

\section{Benchmarks}
\label{sec:benchmarks}

We compare our method's sequence detection performance and speed to two other recently published approaches for which code is freely available: \emph{seqNMF}\footnote{https://github.com/FeeLab/seqNMF} and \emph{PP-Seq}\footnote{https://github.com/lindermanlab/PPSeq.jl}. 

\subsection{Detection performance}
\label{sub:detection_perf}

To evaluate the models' ability to detect sequences, we construct a grid of synthetic datasets of varying difficulty by manipulating (1) pattern sparsity (spike dropout probability), (2) inter-sequence interval (number of time steps between the sequences), (3) sequence length (number of neurons in a sequence before applying dropout), and (4) jitter (standard deviation, in time steps, by which spike timing deviates from its ideal timing). 

\begin{figure}[H] 
\centering
\includegraphics[width=0.48\textwidth]{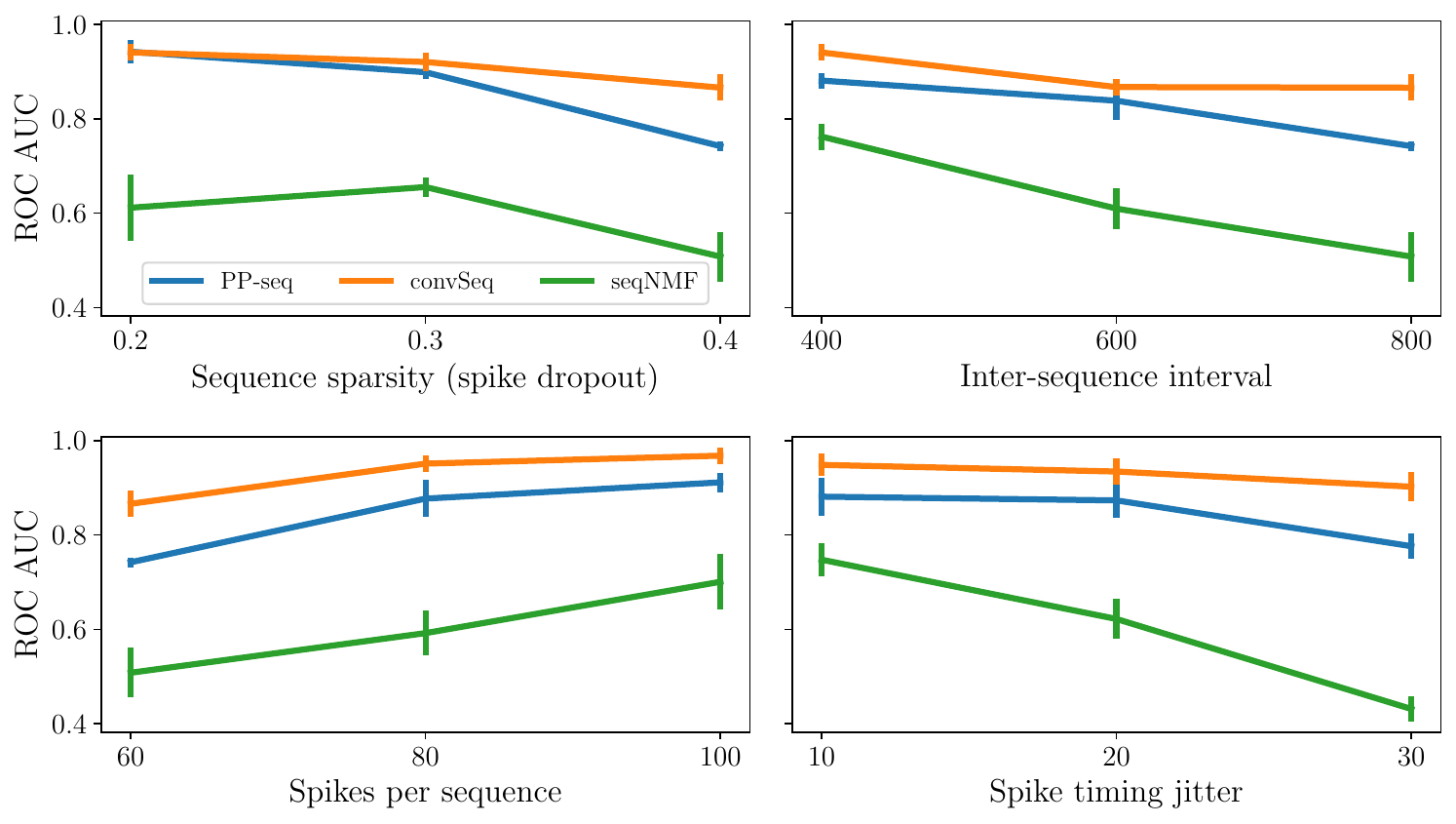}
\vskip -0.1in
\caption{Detection performance of \emph{PP-Seq}, \emph{seqNMF} and \emph{convSeq} (ours) on a grid of 81 datasets. Error bars indicate 95\% confidence intervals over 5 runs on the same dataset.}
\label{fig:ROC_AUC}
\end{figure}

Occurrences of one sequence, with a unique combination of parameters for each dataset, were embedded in the same background activity matrix (452 neurons by 18137 time steps) obtained by permuting the rows and columns of the real recording of the CA1 area of the hippocampus of a mouse. On each of the datasets, models were allowed to run for 100 epochs.

For \emph{PP-Seq} and \emph{seqNMF}, ROC curves were calculated by thresholding the posterior distribution and the factor's temporal loadings, respectively (as in \citet{williams2020point}) and, similarly, the convolution curve $\mathbf{\hat{x}}$ in our method.

Fig. \ref{fig:ROC_AUC}, suggests that, for all the models, the detection performance is genrally better for strong (i.e. involving relatively many neurons), dense (have a relatively low spike dropout probability), temporally constistent (low jitter) and frequent (short inter-sequence interval). However, our method outperforms the baselines in all the conditions tested.

\subsection{Speed}
\label{speed_benchmarks}

We next show how our method's run time scales as a function of dataset size compared to that of the baselines. Using the same hardware, we ran the methods on a grid of datasets, each with the same number of neurons and sequence properties (Appendix \ref{appendix:speed_bench}), but a different number of timesteps, $T$, and intensity of background activity, $S$, defined here as $\frac{1}{NT}\sum{\mathbf{X}}$. Each optimization was run for 100 steps. 100 is the default number of optimization steps in the open-source implementations of \emph{PP-seq} and \emph{seqNMF}. In our model, the same number of optimization steps was sufficient for the $\mathrm{Var}(\mathbf{\hat{x}}^{(k)})$ term in the loss function to reach an approximate plateau, indicating no need for further optimization.

\begin{figure}[H]
\centering
\includegraphics[width=0.48\textwidth]{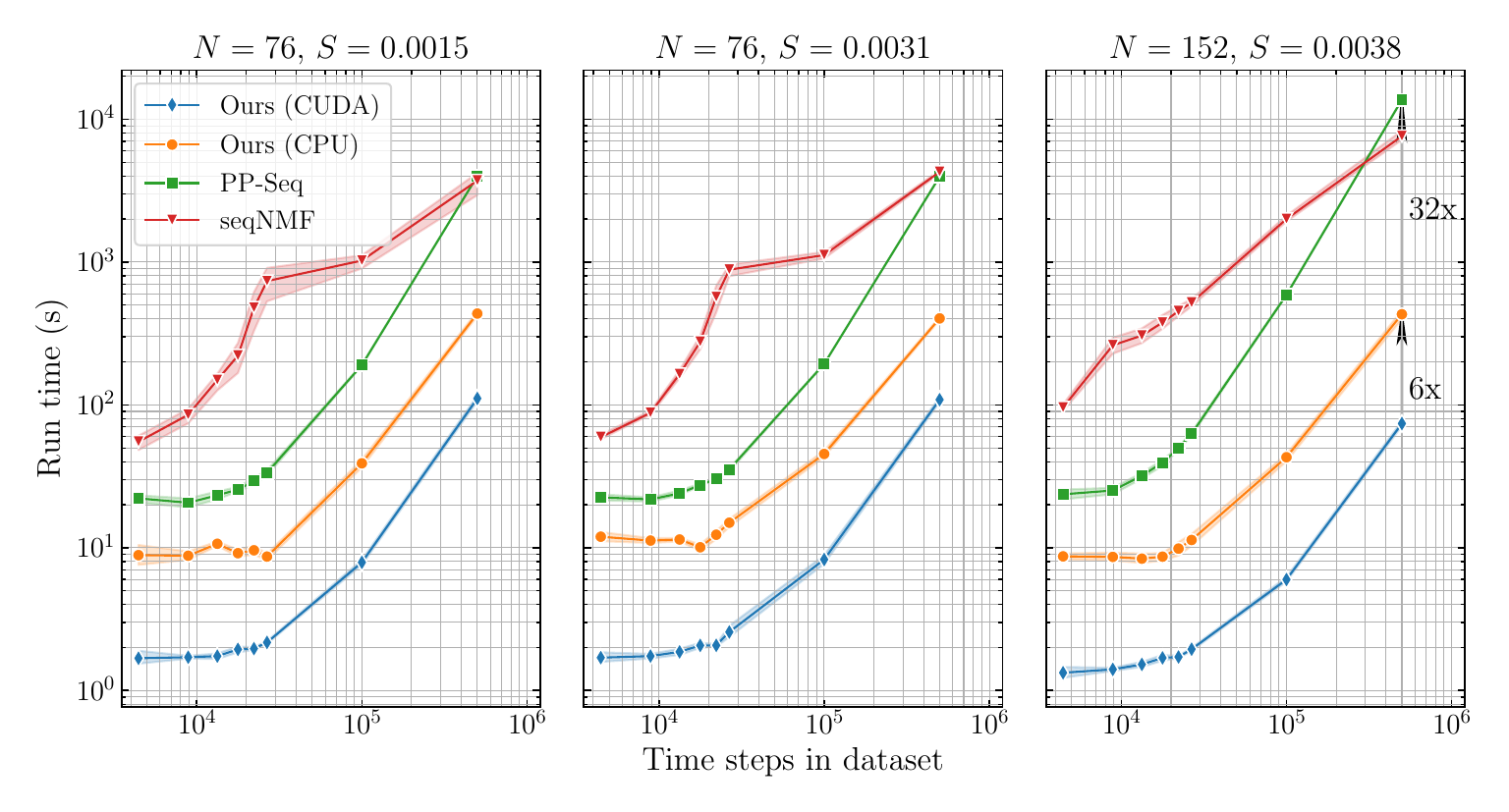}
\vskip -0.1in
\caption{Regardless of the number of neurons ($N$) and intensity of the background activity ($S$), our method outperforms \emph{seqNMF} and \emph{PP-Seq} on the same datasets. Shades indicate 95\% confidence intervals computed over 8 runs.}
\label{fig:speed_benchmarks}
\end{figure}

To ensure as fair a comparison as possible, we first run our method with GPU disabled (orange line in Fig. \ref{fig:speed_benchmarks}). Compared to \emph{PP-Seq}, our approach is about 32 times faster on the largest dataset (500000 timesteps), and enabling the GPU further reduces the run time by a factor of six.

\subsection{Run time scaling as a function of $K$}
\label{appendix:run_time_scaling}

In addition to comparing the run time for one pattern (Fig. \ref{fig:speed_benchmarks}), we compare how the run time scales as a function the assumed number of patterns. Analogously to Section \ref{sub:detection_perf} we constructed synthetic datasets with $N=452$ and $T = 18137$ by embedding $K \in \{1, \dots, 6\}$ sequences (40 neurons each) into the same background activity. The inter-sequence interval and spike dropout were set to 200 timesteps and 0.2, respectively. Each model was fitted for 100 epochs. $M$ was set to 100.

\begin{figure}[H]
\centering
\includegraphics[width=0.48\textwidth]{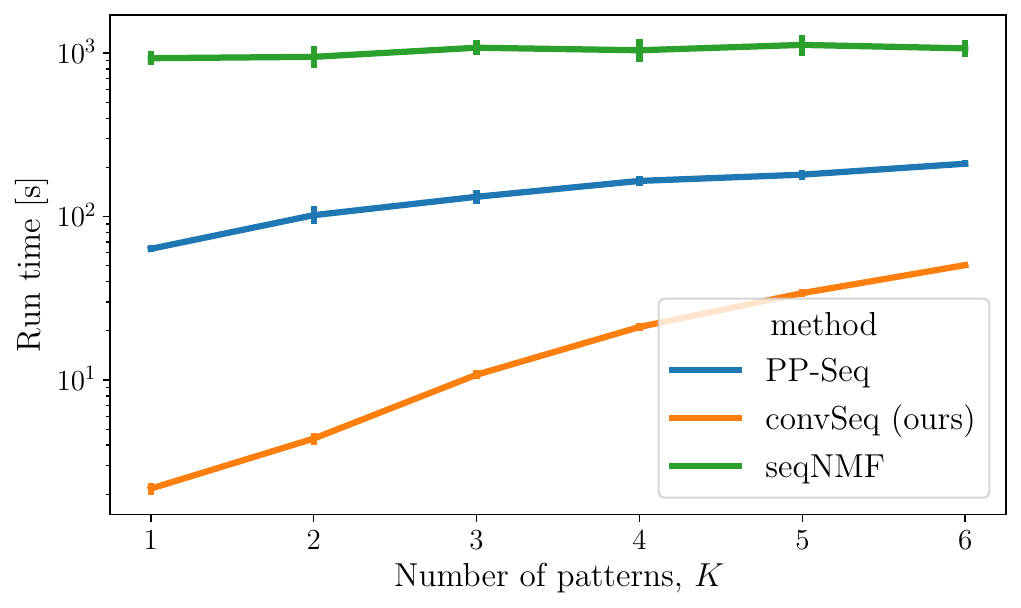}

\vskip -0.1in
\caption{Run time as a function of $K$ (sequence types). Dataset parameters: $T = 18137$, $N = 452$, with one sequence type of 40 neurons with dropout of 0.2, ISI of 200 time steps, and jitter of 10. Error bars indicate the 95\% confidence intervals computed over 8 runs.}
\label{fig:ppseq_vs_our_100_ep}
\end{figure}

Even if $K$ is set to large values, our method still wins compared to the baselines.

\section{Notes on implementation, training and hyperparameter choice}
\label{appendix:guidance}

The model was implemented in Pytorch \citep{NEURIPS2019_9015} and optimized with the Adam \citep{kingma2014adam} optimizer with default parameters except the learning rate which was set to 0.1 for faster convergence. For the 2D convolution operation we used no padding in the dimension of neurons and a padding of $M//2$ zeros in the time dimension to ensure that $\mathbf{\hat{x}}^{(k)}$ has the same number of timesteps as the dataset. The cross-correlation term was implemented as 1D convolution with zero padding of size $M//2$. The total variation term  smoothens the convolution $\mathbf{\hat x}^{(k)}$. We found it to be less important for the second formulation of our method, because the parameterization of $\mathbf{W}^{(k)}$ with truncated Gaussians itself has a strong smoothing effect on the corresponding $\mathbf{\hat x}^{(k)}$ (Appendix \ref{appendix:total_variation}). In general, given the same dataset and filter sizes, one optimization step takes approximately the same time for both formulations of our method. All the experiments were run on a Linux machine with a 64-core AMD EPYC 7702 CPU with 503GB of RAM and an NVIDIA A6000 GPU with 48.67 GB of RAM. We use batch gradient descent, since all the datasets fit entirely into the RAM. However, implementing batched optimization (for even larger datasets or smaller RAM) is straightforward.

\begin{figure}[H]
\centering
\includegraphics[width=0.48\textwidth]{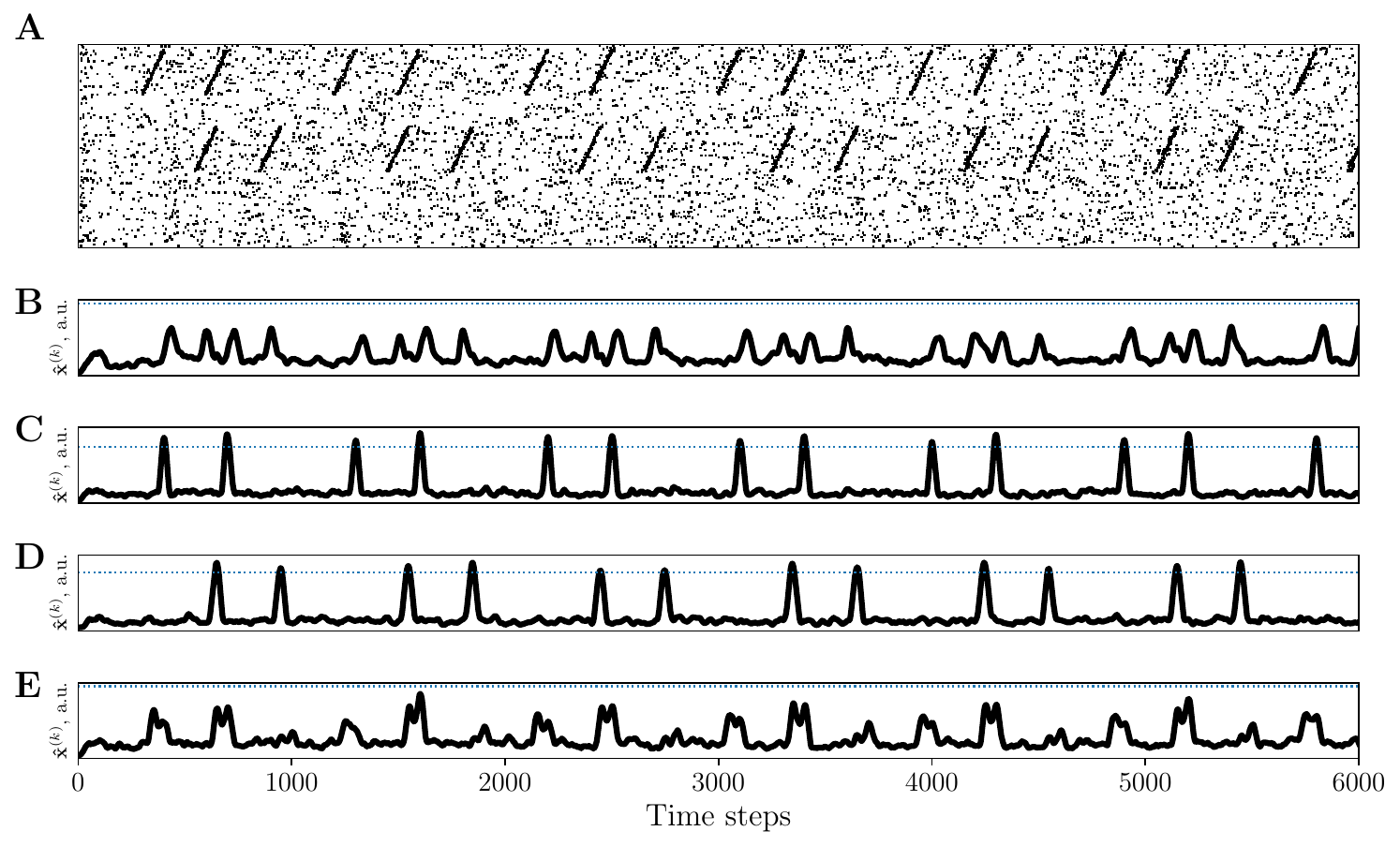}
\vskip -0.1in
\caption{Even with $K=4$, which is greater than the number of sequences (2), and despite partial overlap in time, the second and the third filter (C, D) detect the sequences. The extraneous two filters’ peaks fail to reach the significance threshold (B, E).}
\label{fig:space_overlap}
\end{figure}

\textbf{Choosing $K$.} Similarly to \emph{seqNMF} and \emph{convNMF}, our method still works if the number of filters $K$ is not exactly equal to the actual number of patterns. If $K$ is less than the number of patterns, the filters become tuned to the stronger of the patterns. In the reverse situation, when $K$ happens to be greater than the number of patterns, some “extra” filters’ convolution curves will have a large degree of similarity, but their peaks will not reach statistical significance (e.g. as in Fig \ref{fig:space_overlap} B and E). We found that a good strategy is to start with a conservative choice of $K$ (e. g. $K=1$), and run optimization with progressively larger values of $K$ (such empirical search is realistic owing to the speed of our method). The significance of the convolution peaks provides a good guidance as to whether or not a particular choice of $K$ is good.

Consider the case in which two patterns exist in the data, and one of them is much stronger than the other. With $K=1$ the stronger of the two will be detected. With $K=2$, both patterns will be detected (i.e. the first, already detected, one and the second). Setting $K=3$ should result in still detecting the two patterns plus some spurious “pattern” by the third filter (whose convolution peaks should not reach statistical significance, because the cross-correlation term in Eq. \ref{eq:conv1} penalizes similar activations and, indirectly, filters that are tuned to the same pattern).

\textbf{Choosing $M$.} As with $K$, $M$ should be chosen empirically, unless one has prior knowledge about the length of the expected sequences. It should be noted that

\begin{enumerate}
    \item If $M$ is longer (more than twice the pattern’s length) than the pattern, the share of the spikes participating in the pattern is too small relative to background spikes, effectively reducing the signal-to-noise ratio.
    \item If $M$ is significantly shorter than the pattern (less than half the pattern’s length), the filter might not “see” the pattern in its entirety, which might lead to more than one pattern being tuned to different parts of the same sequence (e.g. one tuned to the beginning and the other tuned to the end of the sequence).
\end{enumerate}

\section{Conclusions and future directions}
\label{section:Conclusions}

In this paper we have proposed a method for unsupervised detection of spatio-temporal patterns in neural recordings which may have practical utility in neuroscience research, especially in situations in which no behavioral references are available. We demonstrated that both on synthetic and real data, our approach is able to detect multiple spike sequences, including those that partially share neurons, or those that involve exactly the same neurons but are expressed in forward and reverse directions. Importantly, our approach is much faster, which unlocks new possibilities for the study of structured spontaneous activity in large-scale neural recordings. In fact, there are at least two ways in which our approach could be made even more efficient (for example, using dilated and/or strided convolutions). Exploring these and other improvements is an interesting direction for future work.

\section*{Acknowledgements}
The first author thanks Ibrahim Alsolami for helpful discussions and acknowledges financial support from KAKENHI grant JP23KJ2131 and Google. We thank anonymous reviewers for their feedback and suggestions. The second author acknowledges support from KAKENHI grant JP23H05476.

\section*{Impact Statement}
\label{section:impact}

This work advances neuroscience by developing a fast and scalable method to detect spatiotemporal patterns in neural activity. Discerning these patterns is crucial for deepening our understanding of cognitive processes in biological brains and could enable breakthroughs in diagnosing and treating mental disorders.

\nocite{langley00}

\bibliography{example_paper}

\begin{thebibliography}{24}
\providecommand{\natexlab}[1]{#1}
\providecommand{\url}[1]{\texttt{#1}}
\expandafter\ifx\csname urlstyle\endcsname\relax
  \providecommand{\doi}[1]{doi: #1}\else
  \providecommand{\doi}{doi: \begingroup \urlstyle{rm}\Url}\fi

\bibitem[Arieli et~al.(1996)Arieli, Sterkin, Grinvald, and Aertsen]{arieli1996dynamics}
Arieli, A., Sterkin, A., Grinvald, A., and Aertsen, A.
\newblock Dynamics of ongoing activity: explanation of the large variability in evoked cortical responses.
\newblock \emph{Science}, 273\penalty0 (5283):\penalty0 1868--1871, 1996.

\bibitem[Beggs \& Plenz(2003)Beggs and Plenz]{beggs2003neuronal}
Beggs, J.~M. and Plenz, D.
\newblock Neuronal avalanches in neocortical circuits.
\newblock \emph{Journal of neuroscience}, 23\penalty0 (35):\penalty0 11167--11177, 2003.

\bibitem[Fiser et~al.(2004)Fiser, Chiu, and Weliky]{fiser2004small}
Fiser, J., Chiu, C., and Weliky, M.
\newblock Small modulation of ongoing cortical dynamics by sensory input during natural vision.
\newblock \emph{Nature}, 431\penalty0 (7008):\penalty0 573--578, 2004.

\bibitem[Foster \& Wilson(2006)Foster and Wilson]{foster2006reverse}
Foster, D.~J. and Wilson, M.~A.
\newblock Reverse replay of behavioural sequences in hippocampal place cells during the awake state.
\newblock \emph{Nature}, 440\penalty0 (7084):\penalty0 680--683, 2006.

\bibitem[Girardeau et~al.(2009)Girardeau, Benchenane, Wiener, Buzs{\'a}ki, and Zugaro]{girardeau2009selective}
Girardeau, G., Benchenane, K., Wiener, S.~I., Buzs{\'a}ki, G., and Zugaro, M.~B.
\newblock Selective suppression of hippocampal ripples impairs spatial memory.
\newblock \emph{Nature neuroscience}, 12\penalty0 (10):\penalty0 1222--1223, 2009.

\bibitem[Grossberger et~al.(2018)Grossberger, Battaglia, and Vinck]{grossberger2018unsupervised}
Grossberger, L., Battaglia, F.~P., and Vinck, M.
\newblock Unsupervised clustering of temporal patterns in high-dimensional neuronal ensembles using a novel dissimilarity measure.
\newblock \emph{PLoS computational biology}, 14\penalty0 (7):\penalty0 e1006283, 2018.

\bibitem[Jutten \& Herault(1991)Jutten and Herault]{JUTTEN19911}
Jutten, C. and Herault, J.
\newblock Blind separation of sources, part i: An adaptive algorithm based on neuromimetic architecture.
\newblock \emph{Signal Processing}, 24\penalty0 (1):\penalty0 1--10, 1991.
\newblock ISSN 0165-1684.
\newblock \doi{https://doi.org/10.1016/0165-1684(91)90079-X}.
\newblock URL \url{https://www.sciencedirect.com/science/article/pii/016516849190079X}.

\bibitem[Kingma \& Ba(2014)Kingma and Ba]{kingma2014adam}
Kingma, D.~P. and Ba, J.
\newblock Adam: A method for stochastic optimization.
\newblock \emph{arXiv preprint arXiv:1412.6980}, 2014.

\bibitem[Langley(2000)]{langley00}
Langley, P.
\newblock Crafting papers on machine learning.
\newblock In Langley, P. (ed.), \emph{Proceedings of the 17th International Conference on Machine Learning (ICML 2000)}, pp.\  1207--1216, Stanford, CA, 2000. Morgan Kaufmann.

\bibitem[Lee \& Wilson(2002)Lee and Wilson]{lee2002memory}
Lee, A.~K. and Wilson, M.~A.
\newblock Memory of sequential experience in the hippocampus during slow wave sleep.
\newblock \emph{Neuron}, 36\penalty0 (6):\penalty0 1183--1194, 2002.

\bibitem[Li et~al.(2022)Li, Qi, and Pan]{li2022online}
Li, W., Qi, Y., and Pan, G.
\newblock Online neural sequence detection with hierarchical dirichlet point process.
\newblock \emph{Advances in Neural Information Processing Systems}, 35:\penalty0 6654--6665, 2022.

\bibitem[Mackevicius et~al.(2019)Mackevicius, Bahle, Williams, Gu, Denisenko, Goldman, and Fee]{mackevicius2019unsupervised}
Mackevicius, E.~L., Bahle, A.~H., Williams, A.~H., Gu, S., Denisenko, N.~I., Goldman, M.~S., and Fee, M.~S.
\newblock Unsupervised discovery of temporal sequences in high-dimensional datasets, with applications to neuroscience.
\newblock \emph{Elife}, 8:\penalty0 e38471, 2019.

\bibitem[Paszke et~al.(2019)Paszke, Gross, Massa, Lerer, Bradbury, Chanan, Killeen, Lin, Gimelshein, Antiga, Desmaison, Kopf, Yang, DeVito, Raison, Tejani, Chilamkurthy, Steiner, Fang, Bai, and Chintala]{NEURIPS2019_9015}
Paszke, A., Gross, S., Massa, F., Lerer, A., Bradbury, J., Chanan, G., Killeen, T., Lin, Z., Gimelshein, N., Antiga, L., Desmaison, A., Kopf, A., Yang, E., DeVito, Z., Raison, M., Tejani, A., Chilamkurthy, S., Steiner, B., Fang, L., Bai, J., and Chintala, S.
\newblock Pytorch: An imperative style, high-performance deep learning library.
\newblock In \emph{Advances in Neural Information Processing Systems 32}, pp.\  8024--8035. Curran Associates, Inc., 2019.

\bibitem[Peter et~al.(2016)Peter, Durstewitz, Diego, and Hamprecht]{peter2016sparse}
Peter, S., Durstewitz, D., Diego, F., and Hamprecht, F.~A.
\newblock Sparse convolutional coding for neuronal ensemble identification.
\newblock \emph{arXiv preprint arXiv:1606.07029}, 2016.

\bibitem[Pfeiffer \& Foster(2013)Pfeiffer and Foster]{pfeiffer2013hippocampal}
Pfeiffer, B.~E. and Foster, D.~J.
\newblock Hippocampal place-cell sequences depict future paths to remembered goals.
\newblock \emph{Nature}, 497\penalty0 (7447):\penalty0 74--79, 2013.

\bibitem[Rubin et~al.(2019)Rubin, Sheintuch, Brande-Eilat, Pinchasof, Rechavi, Geva, and Ziv]{rubin2019revealing}
Rubin, A., Sheintuch, L., Brande-Eilat, N., Pinchasof, O., Rechavi, Y., Geva, N., and Ziv, Y.
\newblock Revealing neural correlates of behavior without behavioral measurements.
\newblock \emph{Nature communications}, 10\penalty0 (1):\penalty0 4745, 2019.

\bibitem[Schrader et~al.(2008)Schrader, Grun, Diesmann, and Gerstein]{schrader2008detecting}
Schrader, S., Grun, S., Diesmann, M., and Gerstein, G.~L.
\newblock Detecting synfire chain activity using massively parallel spike train recording.
\newblock \emph{Journal of neurophysiology}, 100\penalty0 (4):\penalty0 2165--2176, 2008.

\bibitem[Shimazaki et~al.(2012)Shimazaki, Amari, Brown, and Gr{\"u}n]{shimazaki2012state}
Shimazaki, H., Amari, S.-i., Brown, E.~N., and Gr{\"u}n, S.
\newblock State-space analysis of time-varying higher-order spike correlation for multiple neural spike train data.
\newblock \emph{PLoS computational biology}, 8\penalty0 (3):\penalty0 e1002385, 2012.

\bibitem[Smaragdis(2004)]{smaragdis2004non}
Smaragdis, P.
\newblock Non-negative matrix factor deconvolution; extraction of multiple sound sources from monophonic inputs.
\newblock In \emph{Independent Component Analysis and Blind Signal Separation: Fifth International Conference, ICA 2004, Granada, Spain, September 22-24, 2004. Proceedings 5}, pp.\  494--499. Springer, 2004.

\bibitem[Smaragdis(2006)]{smaragdis2006convolutive}
Smaragdis, P.
\newblock Convolutive speech bases and their application to supervised speech separation.
\newblock \emph{IEEE Transactions on Audio, Speech, and Language Processing}, 15\penalty0 (1):\penalty0 1--12, 2006.

\bibitem[Stringer et~al.(2023)Stringer, Zhong, Syeda, Du, Kesa, and Pachitariu]{Stringer2023.07.25.550571}
Stringer, C., Zhong, L., Syeda, A., Du, F., Kesa, M., and Pachitariu, M.
\newblock Rastermap: a discovery method for neural population recordings.
\newblock \emph{bioRxiv}, 2023.
\newblock \doi{10.1101/2023.07.25.550571}.
\newblock URL \url{https://www.biorxiv.org/content/early/2023/08/07/2023.07.25.550571}.

\bibitem[Torre et~al.(2016)Torre, Canova, Denker, Gerstein, Helias, and Gr{\"u}n]{torre2016asset}
Torre, E., Canova, C., Denker, M., Gerstein, G., Helias, M., and Gr{\"u}n, S.
\newblock Asset: analysis of sequences of synchronous events in massively parallel spike trains.
\newblock \emph{PLoS computational biology}, 12\penalty0 (7):\penalty0 e1004939, 2016.

\bibitem[Watanabe et~al.(2019)Watanabe, Haga, Tatsuno, Euston, and Fukai]{watanabe2019unsupervised}
Watanabe, K., Haga, T., Tatsuno, M., Euston, D.~R., and Fukai, T.
\newblock Unsupervised detection of cell-assembly sequences by similarity-based clustering.
\newblock \emph{Frontiers in Neuroinformatics}, pp.\ ~39, 2019.

\bibitem[Williams et~al.(2020)Williams, Degleris, Wang, and Linderman]{williams2020point}
Williams, A., Degleris, A., Wang, Y., and Linderman, S.
\newblock Point process models for sequence detection in high-dimensional neural spike trains.
\newblock \emph{Advances in neural information processing systems}, 33:\penalty0 14350--14361, 2020.

\end{thebibliography}
\bibliographystyle{icml2024}

\onecolumn
\newpage

\twocolumn
\appendix
\section*{Appendix}
\section{Hyperparameters}
\label{HPs}

\begin{table}[ht!]
  \caption{Hyperparameters}
  \label{table:HPs}
  \centering
  \begin{tabular}{p{0.15\columnwidth} p{0.4\columnwidth} p{0.3\columnwidth}}
    \toprule
    HP & Description & Value \\
    \midrule
    $\beta_{\mathrm {xcor}}$ & Diversity loss weight & 0 if $K=1$ else 10.0 ($10^5$ in Fig. \ref{fig:Gauss_Fig7})     \\
    $\beta_{\mathrm {TV}}$   & Total variation weight & 15.2 in Fig. \ref{fig:fig1}, 4.5 in Fig. \ref{fig:speed_benchmarks}, otherwise 100.0   \\
    $M$ & Filter width (along the time dimension) & 200 in Figs. \ref{fig:fig1} and \ref{fig:fig9}, otherwise 100 \\
    \verb+lrate+     & Learning rate   & 0.1 \\
    $j$     & Maximum cross-correlation distance   & $M$         \\
    $\sigma$     & Standard deviation of the filters' Gaussians (2nd formulation) & 16.0 (20.0 in Fig. \ref{fig:Gauss_Fig7}        \\
   &          \\
    \bottomrule
  \end{tabular}
\end{table}

In general, the total variation term can be set to zero in the formulation with truncated Gaussians (see main text), especially with relatively large values of $\sigma$. In cases that in involve overlapping sequences (as in Figs. \ref{fig:fig5}, \ref{fig:fig6} and \ref{fig:fig9}). We have also observed the need for a large weight for the cross-correlation penalty in Eq. \ref{eq:conv1}

\section{Algorithms}
\label{appendix:algo1}

\begin{algorithm}
\caption{With minimally constrained filters}
\label{alg:cap0}
\begin{algorithmic}

\STATE \textbf{Input:} \textbf{X}, $K$, steps
\FOR {$k \in \{1,..., K\}$}
    \STATE Initialize $\textbf{W}^{(k)}$
\ENDFOR

\FOR {steps}
    \STATE Take a gradient step for $\mathcal{L}$
    \STATE Update $\textbf{W}^{(k)}$, $k \in \{1,..., K\}$
\ENDFOR
\FOR {$k \in \{1,..., K\}$}
    \STATE Sort the rows of $\textbf{W}^{(k)}$ according to the latency of the
    \STATE     maximum within-row value, record sorting indices $\textbf{s}$ of size $N$
    \STATE Obtain $\textbf{X}^{(k)}$ by re-ordering the rows of $\textbf{X}$ with $\textbf{s}$
\ENDFOR
\end{algorithmic}
\end{algorithm}

\begin{algorithm}
\caption{With parameterized truncated Gaussians}
\label{alg:cap1}
\begin{algorithmic}

\STATE \textbf{Input:} \textbf{X}, $K$, steps, $\sigma$
\FOR {$k \in \{1,..., K\}$}
    \FOR {$n \in \{1,..., N\}$}
        \STATE Make a truncated Gaussian $\textbf{g}^{(k)}_{n}$ with mean $\mu^{(k)}_{n} \sim \mathcal{U}(1,M)$ and standard deviation $\sigma$
        \STATE Set the $n$-th row of $\textbf{W}^{(k)}$ equal to $\textbf{g}^{(k)}_{n}$
    \ENDFOR
\ENDFOR

\FOR {steps}
    \STATE Take a gradient step for $\mathcal{L}$
    \STATE Update $\mu^{(k)}_{n}$, $k \in \{1,..., K\}$
    \STATE Construct a new $\textbf{W}^{(k)}$, whose rows are truncated Gaussians with $\mu^{(k)}_{n}$,  $k \in \{1,..., K\}$
\ENDFOR

\FOR {$k \in \{1,...,K\}$}
    \STATE Sort $\boldsymbol{\mu}^{(k)}$, record sorting indices $\textbf{s}$
    \STATE Obtain $\textbf{X}^{(k)}$ by re-ordering the rows of $\textbf{X}$ with $\textbf{s}$
\ENDFOR
\end{algorithmic}
\end{algorithm}

\section{Dataset and sequence properties used for speed benchmarks}
\label{appendix:speed_bench}
Each dataset of $N \in \{76, 152\}$ neurons was constructed out of background activity matrices with $T \in \{4441, 8882, 13323, 17764, 22205, 26646, 100000, 500000\}$ timesteps and with background spiking intensity $S \in \{0.0015, 0.0031, 0.0038\}$. Into these background activity matrices we embedded sequences of 40 neurons, each with the following fixed parameters: dropout probability of 0.2, inter-sequence interval of 200 timesteps, and the standard deviation of spike timing (jitter) of 10 timesteps.

\section{Total variation}
\label{appendix:total_variation}

\begin{figure}[H]
\centering
\includegraphics[width=0.48\textwidth]{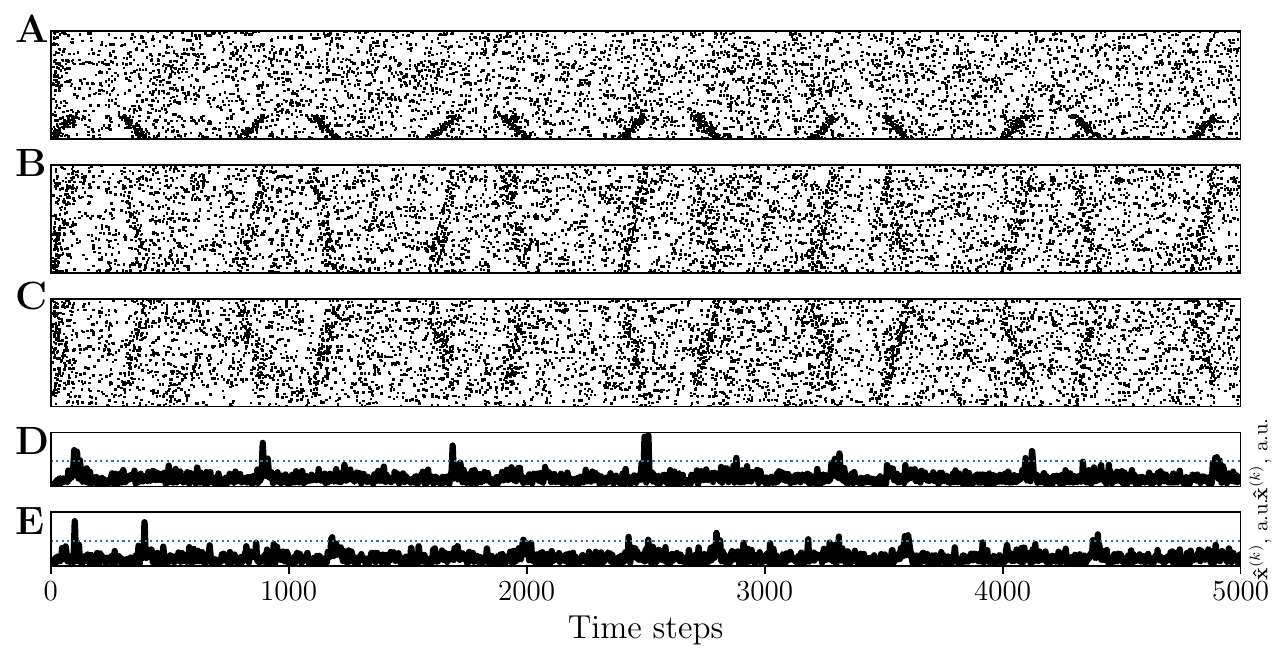}
\vskip -0.1in
\caption{With $\beta_{TV} = 1.5$, the model produces false positives. Panels D and E show the response of the first and second filters, respectively.}
\label{fig:appendix_TV1}
\end{figure}

\begin{figure}[H]
\centering
\includegraphics[width=0.48\textwidth]{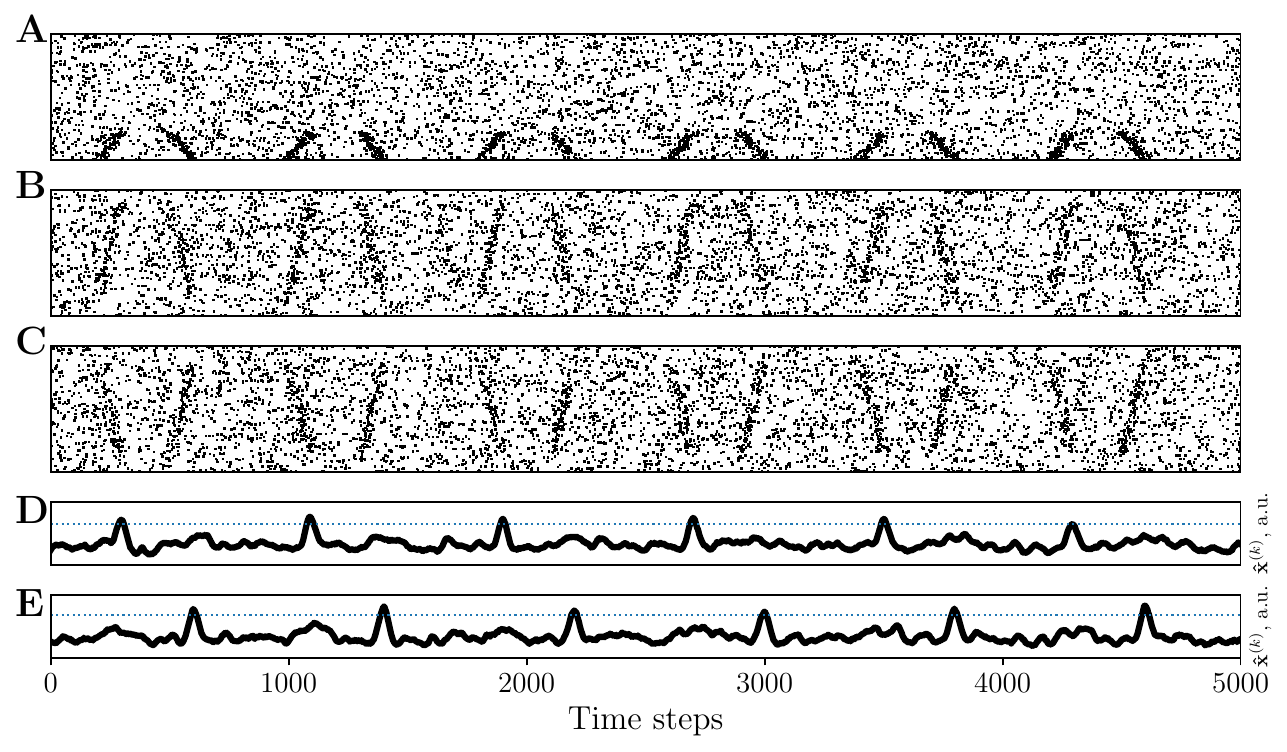}
\vskip -0.1in
\caption{With $\beta_{TV} = 100$, no false positives are present, the filters' responses (panels D and E) are smooth and easy to interpret.}
\label{fig:appendix_TV2}
\end{figure}

The total variation term encourages convergence to smooth $\mathbf{\hat x}^{(k)}$. We found that insufficient values of $\beta_{TV}$ increase the likelihood of a false positive (compare Fig. \ref{fig:appendix_TV1} and Fig. \ref{fig:appendix_TV2}). 

\section{Results for the second formulation of the method}
\label{appendix:2nd_additional}

The results reported in the main text were generated using the first formulation of our method. To illustrate that the second method performs comparably, here we provide figures generated using the second formulation.

\begin{figure}[H]
\centering
\includegraphics[width=0.48\textwidth]{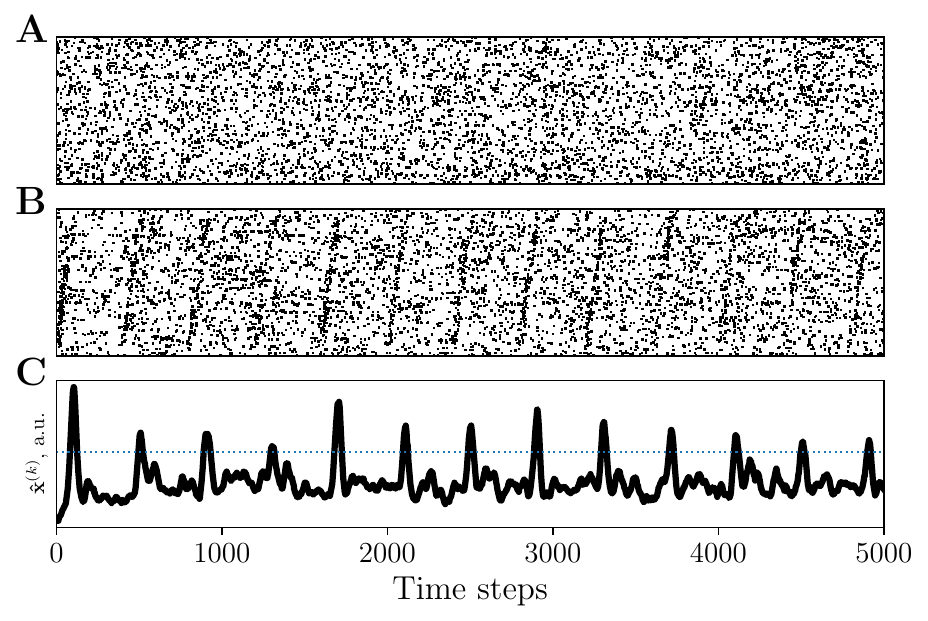}
\vskip -0.1in
\caption{Same as Fig. \ref{fig:fig1} in the main text.}
\label{fig:Gauss_Fig1}
\end{figure}

\begin{figure}[H]
\centering
\includegraphics[width=0.48\textwidth]{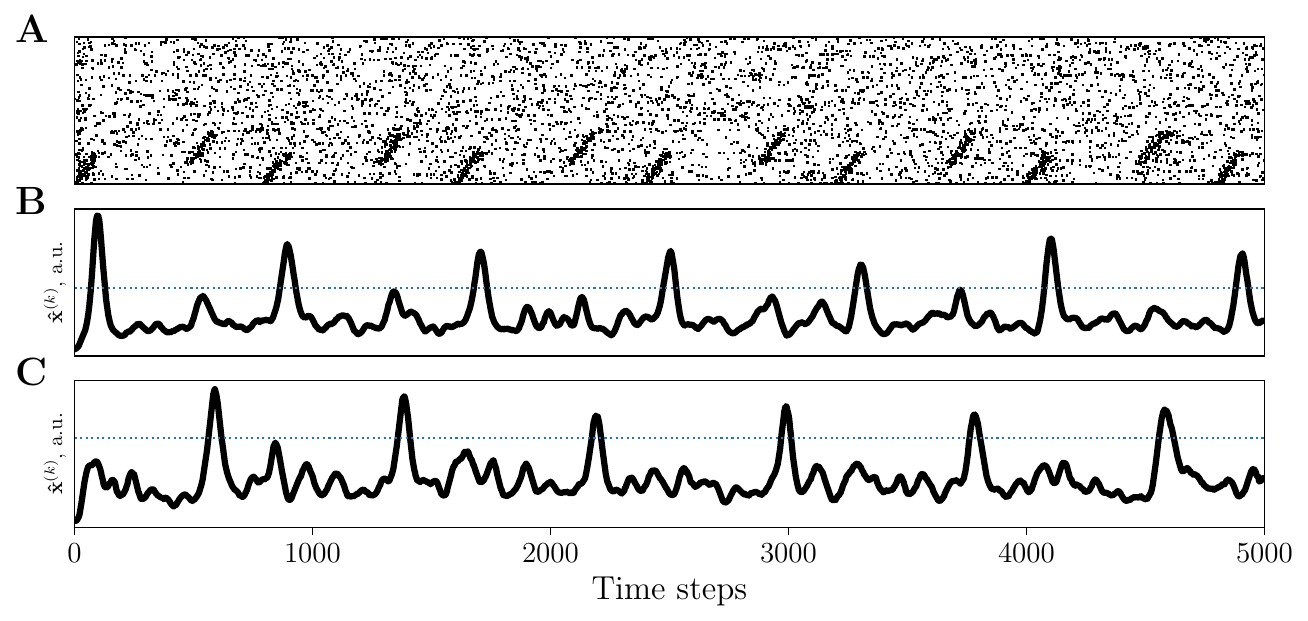}
\vskip -0.1in
\caption{Same as Fig. \ref{fig:fig5} in the main text.}
\label{fig:Gauss_Fig5}
\end{figure}

\begin{figure}[H]
\centering
\includegraphics[width=0.48\textwidth]{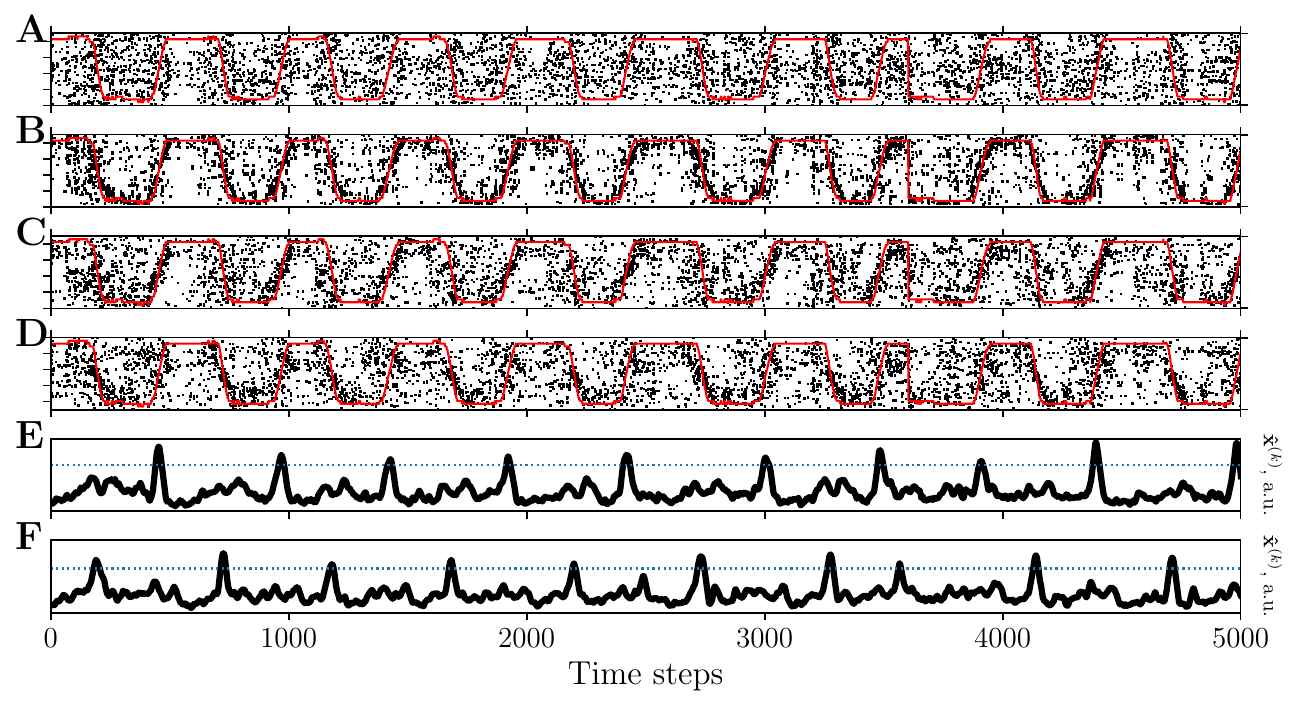}
\vskip -0.1in
\caption{Same as Fig. \ref{fig:fig9} in the main text.}
\label{fig:Gauss_Fig7}
\end{figure}

\end{document}